%% file: main.tex
\documentclass[journal]{IEEEtran}

\usepackage{graphicx}
\usepackage{amsmath}
\usepackage{tikz}
\usepackage{pgfplots}
\usepackage{mathtools}
\usepgfplotslibrary{colorbrewer}
\usepgfplotslibrary{groupplots}
\usetikzlibrary{shapes.geometric}
\usepackage{pgfplotstable}
\pgfplotsset{compat=1.17}
\usetikzlibrary{backgrounds}
\usepackage{wrapfig}  
\usepackage{subfigure}
\usepackage{amsfonts}
\usepackage{multirow}
\usetikzlibrary{arrows}
\usetikzlibrary{decorations.markings}
\usetikzlibrary{matrix}
\usepackage{xcolor}
\usepackage{colortbl}
\usepackage{threeparttable}
\usepackage{environ}
\usepackage{multicol}
\usepackage[noadjust]{cite}

\newcommand{\NLI}{\ensuremath{\text{NLI}}}
\newcommand{\SNR}{\ensuremath{\mathrm{SNR}}}

\newcommand{\ts}{\textsuperscript}

\DeclareMathOperator{\asinh}{asinh} 
\DeclareMathOperator{\atan}{atan}
\DeclareMathOperator{\sign}{sign}

\NewEnviron{splitfit}[1][\linewidth]{%
\resizebox{#1}{!}{$\begin{aligned}\BODY\end{aligned}$}%
}

\makeatletter
\newcommand{\biggg}{\bBigg@{2.5}}
\newcommand{\Biggg}{\bBigg@{3}}
\makeatother
\def\bigggl{\mathopen\biggg}
\def\bigggr{\mathclose\biggg}
\def\Bigggl{\mathopen\Biggg}
\def\Bigggr{\mathclose\Biggg}

\begin{document}

\title{A Closed-form Expression for the Gaussian Noise Model in the Presence of Raman Amplification}

\author{H.~Buglia, M. Jarmolovi\v{c}ius, L.~Galdino, R.I. Killey,
        and~P.~Bayvel
\thanks{This work is partly funded by the EPSRC Programme Grants TRANSNET (EP/R035342/1) and EWOC (EP/W015714/1).  H. Buglia and M. Jarmolovi\v{c}ius are funded by an EPSRC studentship (EP/T517793/1), the Microsoft 'Optics for the Cloud' Alliance and a UCL Faculty of Engineering Sciences Studentship.}
\thanks{The authors are with the Optical Networks Group, University College London, Department of Electronic and Electrical Engineering, Roberts Building, Torrington Place, London
WC1E 7JE, U.K. L. Galdino is with Corning Optical Communications, Ewloe CH5 3XD, U.K (e-mail: \{henrique.buglia.20; min.jarmolovicius.17; r.killey; p.bayvel;\}@ucl.ac.uk, galdinol@corning.com).}
}

\maketitle

\begin{abstract}
A closed-form model for the nonlinear interference (NLI) in Raman amplified links is presented, the formula accounts for both forward (FW) and backward (BW) pumping schemes and inter-channel stimulated Raman scattering (ISRS) effect. The formula also accounts for an arbitrary number of pumps, wavelength-dependent fibre parameters, launch-power profiles, and is tested over a distributed Raman-amplified system setup. The formula is suitable for
ultra-wideband (UWB) optical transmission systems and is applied in a signal with 13~THz optical bandwidth corresponding to transmission over
the S-, C-, and L- band. The accuracy of the closed-form formula is validated through comparison with numerical
integration of the Gaussian noise (GN) model and split-step Fourier method
(SSFM) simulations in a point-to-point transmission link.
\end{abstract}


\IEEEpeerreviewmaketitle

\section{Introduction}

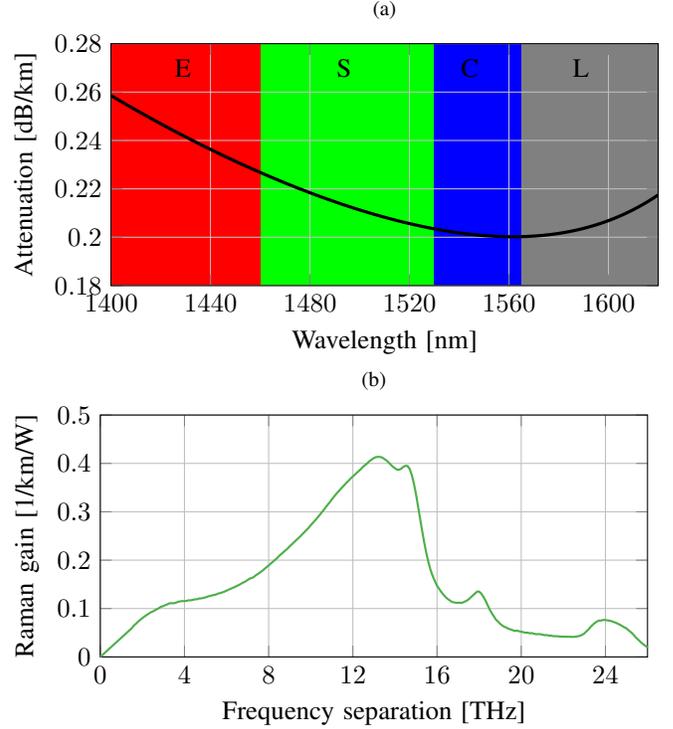
\begin{figure}[ht!]
 \centering
  \input{Figures/loss_and_raman}
\caption{Attenuation coefficient (a) and Raman gain spectrum (b) of an ITU-T G652.D fibre.}\label{fig:raman_loss}
\end{figure}

\IEEEPARstart{T}{o} cope with the exponential growth of data transmission required by internet services such as high-definition video streaming, cloud computing, artificial intelligence, Big Data and the Internet of Things, new technologies such as UWB transmission and space-division multiplexing (SDM) have been widely explored in recent years~\cite{UWBsysm,challenges,bugliaJOCN}. For UWB transmission systems, exploring the low-loss wavelength window of a
silica-based optical fibre, as shown in Fig.~\ref{fig:raman_loss}, requires the utilisation of new amplifier technologies in addition to Erbium-doped fibre amplifiers (EDFAs). Among these, we can cite Thulium and bismuth-doped fibre amplifiers (TDFAs and BDFAs), semiconductor optical
amplifiers (SOA) and Raman amplifiers.

Recently, a wide range of works has shown the benefits of using Raman amplification (RA) to achieve higher throughputs~\cite{Souza:22,Puttnam:22,Puttnam:21,NTT,galdino2020optical,Charlet:22,107Tbps,JLTLidia,Maria,longhaul}. RA can be divided into two types, namely distributed RA and discrete RA. For the former, the pumps are injected into the transmission fibre, while for the latter a separate fibre is used as the amplification stage. In both cases, the pumps interact with the signal to provide the desired signal amplification. 

Together with new amplifier technologies, the key goals in optical network design are to maximise system throughput and introduce intelligence in the network, delivering capacity when and where it is needed~\cite{Matzner:21,Luo:23}. To that purpose, real-time estimation of the UWB system performance is essential, as it enables efficient and rapid system design, online network optimisation routines and virtualisation of the physical layer. 

Such real-time prediction of UWB optical fibre transmission systems can be achieved via closed-form expressions of the GN model and its extensions~\cite{gnmodel,Carena:14,isrsgnmodel}. This model offers a simple way of estimating the fibre NLI by treating it as additive Gaussian noise. Numerous closed-form expressions have been proposed to date~\cite{boscoOFC}. Of interest for UWB transmission systems are closed-form expressions for the GN model in the presence of ISRS effect~\cite{isrsgnmodel}, namely ISRS GN model. Closed-form expressions of this model were derived in~\cite{generalizedclosed,danielclosed,closedzefre,ISRSGNmodel_correction,zefreh2020realtime,MZraman,poggiolini_closed_ecoc,bugliaECOC,bugliaJLT}.    

This work focuses on the derivation of a closed-form formula to estimate the NLI in Raman-amplified links. Apart from~\cite{zefreh2020realtime,MZraman}, the remaining closed-form expressions are valid or tested for lumped-amplified links only. Despite the closed-form formula in~\cite{zefreh2020realtime,MZraman} being valid and tested for Raman amplified links, it is limited to FW pumping schemes and was tested only over C-band systems. A closed-form formula limited to BW pumping schemes can be found in~\cite{RamanDan}, however, it is only valid for C-band systems and limited to 2\ts{nd}~order Raman amplification, i.e., the utilisation of two or fewer pumps. 

In this work, we developed a general closed-form expression of the ISRS GN model~\cite{isrsgnmodel} supporting both FW-RA and BW-RA, ISRS, valid for arbitrary-order RA, i.e., an arbitrary number of pumps. This was enabled by deriving for the first time a semi-analytical solution to model the signal profile in the presence of RA and ISRS. The proposed closed-form formulation is valid for Gaussian constellations, and in this work is tested using a distributed RA setup. Its accuracy is verified with numerical integration of the ISRS~GN model and SSFM simulations.  

The closed-form expression presented in this work was first published in~\cite{bugliaOFC}. In this work, we extensively discuss its validation and present all the mathematical derivations used to obtain it. We also include a complete discussion on the semi-analytical approach used to obtain an accurate estimation of the fibre signal profile evolution along the fibre distance. This work together with~\cite{bugliaOFC} represents the first closed-form expression of the GN model supporting FW-RA and BW-RA in the presence of ISRS.

\section{The signal profile evolution}

This section shows the derivation of the semi-analytical expression for the signal power evolution along the fibre distance in the presence of RA and ISRS. The second part of this section shows the accuracy of the utilisation of the proposed approach.

\subsection{The derivation of the closed-form expression for signal profile evolution}
\label{sec:signal_profile}

For NLI estimation expressions based on regular perturbation analysis, such as the GN model and its extensions~\cite{gnmodel,isrsgnmodel,Carena:14}, the estimation of the NLI interference is dependent on the signal power profile evolution along the optical fibre distance. Because of this, a fundamental step in deriving any closed-form expression for NLI estimation is to first derive a closed-form expression for the signal power profile evolution.



\begin{table*}[bt]
\caption{Pumps' power and wavelength allocation which yields the power profiles shown in Fig~\ref{fig:profile}.}
\centering
\label{tab:pumps}
\input{Figures/pumps.tex}

\end{table*}


In the case of C-band systems, such an expression is trivial as the signal power evolution is only loss dependent~\cite{gnmodel}. The situation is more tricky in the presence of ISRS as the power of each channel interacts with one another and a set of coupled differential equations must be solved. Analytical expressions for this case were derived in~\cite{raman_analytical,zirngibl1998analytical}. These expressions are used in~\cite{danielclosed} (Eq. 16 and 17) to derive a semi-analytical solution of the signal power profile evolution. The solution is semi-analytical because it is further optimised to correctly reproduce the solution of the coupled differential equations.

\begin{figure}[b!]
\input{Figures/power_profile.tex}
\caption{Per-channel launch power evolution along the fibre distance for (a) FW-RA  (b) BW-RA and (c) FW+BW-RA.}
\label{fig:profile}
\end{figure}
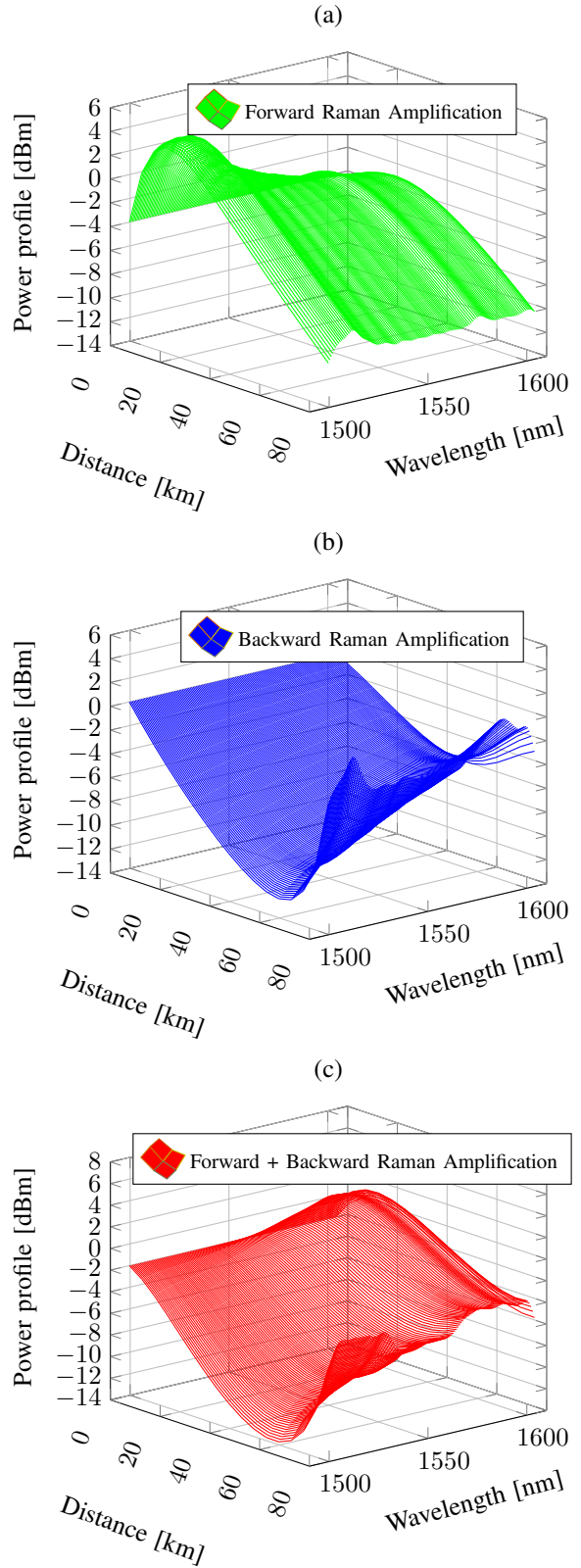

The situation is even more complicated in the case of RA, whereas besides the channel-channel interactions, pump-signal and pump-pump interactions must also be considered, not only in the forward direction but also in the backward one. Indeed, for the case of RA and ISRS, the solution of the so-called coupled differential Raman equations in the presence of RA must be solved, and it is given by

\begin{equation}
\begin{aligned}
\pm  &\frac{\partial P_i}{\partial z} =  - \sum_{k=i+1}^{\text{N}_{\text{ch}}} \frac{f_k}{f_i} g(\Delta f) P_k P_i - \sum_{p:f_i>f_p} \frac{f_p}{f_i} g(\Delta f) P_p P_i + \\
  & + \sum_{k=1}^{i-1} g(\Delta f) P_k P_i + \sum_{p:f_i<f_p} g(\Delta f) P_p P_i - \alpha_i P_i,
\end{aligned}
\label{eq:diff_Raman}
\end{equation}
where, $P_i$, $f_i$ are the power and frequency of the channel of interest (COI), $P_k$, $f_k$ are the power and frequency of the remaining WDM channels, $P_p$, $f_p$ are the power and the frequency of the pumps, $g_r(\Delta f)$ is the polarization averaged, normalized (by the effective core area $A_\text{eff}$) Raman gain spectrum for a frequency separation $\Delta f = |f_i - f_k|$, $j=k,p$ and $\alpha_i$ is the frequency-dependent attenuation coefficient. Note that the symbol $\pm$ represents the pump under consideration, i.e., $+$ for FW-pump and $-$ for BW-pump configurations. The pump equations are obtained by replacing $i=p$ in Eq.~\eqref{eq:diff_Raman}.

The first step in deriving the proposed closed-form expression for NLI estimation in this paper is to find a semi-analytical expression for Eq.~\eqref{eq:diff_Raman}. Semi-analytical approaches were used 
in~\cite{MLfitting, danielclosed, MZraman, RamanDan} to model specific transmission setups. However, other types of approaches are also possible, e.g.~\cite{raman_serena}. In this paper for the first time, we proposed a general semi-analytical solution to account for any RA setup scenario with ISRS effect. 

To carry this derivation, we based ourselves in~\cite{raman_analytical} and show the derivation in Appendix~\ref{appA:triangular_approximation}. Let $\rho(z,f_i)$ be the signal profile evolution normalised by the input power profile, i.e., $\rho(z,f_i) = \frac{P(z,f_i)}{P(0,f_i)}$. Thus, a semi-analytical solution of Eq.~\eqref{eq:diff_Raman} is given by
\begin{equation}\label{eq:Raman_taylor}
\rho(z,f_i) = e^{-\alpha_i z} [1 - (C_{f,i} P_{f}L_\text{eff} + C_{b,i} P_{b} \tilde{L}_\text{eff})(f_i - \hat{f})],
\end{equation}
where 
\begin{align*}
    L_\text{eff}(z) &= (1-e^{-\alpha_{f,i} z})/\alpha_{f,i}\quad,\\
\tilde{L}_\text{eff}(z) &= (e^{-\alpha_{b,i}(L-z)}-e^{-\alpha_{b,i} L})/\alpha_{b,i}\quad,
\end{align*}
$L$ is the span length, $\alpha_i$, $\alpha_{f,i}$ and $\alpha_{b,i}$ are fibre attenuation coefficients, $\hat{f}$ is the average frequency of the FW and BW pumps, $P_{f}$, and $P_{b}$ are the total launch power respectively from the WDM channels together with any FW pumps, and the BW pumps, $C_{f,i}$ and $C_{b,i}$ is the slope of a linear regression of the normalised Raman gain spectrum. The proof of Eq.~\eqref{eq:Raman_taylor} is given in Appendix~\ref{appA:triangular_approximation}.

The coefficients $\alpha_i$, $C_{f,i}$, $C_{b,i}$, $\alpha_{f,i}$, and $\alpha_{b,i}$ are channel-dependent parameters and matched using nonlinear least-squares fitting to correctly reproduce the solution of the Raman differential equations in the presence of RA, which is obtained by numerically solving Eq.~\eqref{eq:diff_Raman}. Note that, three different loss coefficients ($\alpha_i$, $\alpha_{f,i}$, and $\alpha_{b,i}$) and two different slopes of the Raman gain spectrum ($C_{f,i}$ and $C_{b,i}$) are considered - this enables an increase in the dimension of optimisation space and is essential for modelling all the RA scenarios. The parameters $\alpha_i$, $C_{f,i}$, $C_{b,i}$, $\alpha_{f,i}$, and $\alpha_{b,i}$ can be interpreted as modelling respectively the fibre loss, the gain/loss due to FW-RA and BW-RA together with ISRS and how fast the channel gain/loss due to the FW-RA and BW-RA together with ISRS extinguishes along the fibre. This fitting optimisation overcomes the restrictive assumptions used to derive Eq.~\eqref{eq:Raman_taylor} and enables its utilisation in any simulation scenario, such as any number of pumps, launch power profiles and bandwidths.

A main difference between the semi-analytical approach proposed here and the one in~\cite{danielclosed}, is the utilisation of 5 optimisation coefficients, against 3 for the latter. The 2 additional coefficients are essential to model BW-RA.
Note that, our approach is valid for arbitrary-order RA, i.e., an arbitrary number of Raman pumps. The approach is also a generalisation of~\cite{danielclosed} as it is also valid for lumped amplification - if one sets $C_{b,i} = 0$ and $\hat{f} = 0$, the semi-analytical solution for the normalised signal profile shown in~\cite{danielclosed} is obtained.

\subsection{Results for signal profile evolution estimation}
\label{sec:results_signal_profile}

This section illustrates the utilisation of the semi-analytical solution proposed in Eq.~\eqref{eq:Raman_taylor} to reproduce the solution of the differential Raman equations in Eq.~\eqref{eq:diff_Raman}.

The transmission setup consists of a WDM signal with $N_{\text{ch}}$=131 channels spaced by 100~GHz and centred at 1550~nm. The signal is amplified using distributed RA. Each channel was modulated at the symbol rate of 96~GBd, resulting in a total bandwidth of 13~THz~(105~nm), ranging from 1500~nm to 1605~nm, corresponding to the transmission over the S- (1470~nm~-~1530nm), C- (1530~nm~-~1565nm) and L- (1565~nm~-~1615nm) bands. Gaussian symbols are considered in the transmission.
For both scenarios, the span length is 80~km and an ITU-T G652.D fibre is considered with attenuation profile and the Raman gain spectrum shown in Fig.~\ref{fig:raman_loss}. 

We consider three different simulation scenarios: FW-RA, BW-RA and FW+BW-RA. A spectrally uniform launch power profile, where each channel carries -4~dBm, 0~dBm and -2~dBm is considered respectively for each one of the scenarios. For all the cases, the number of pumps, and their wavelengths and powers are chosen based on a "find minimum of constrained nonlinear multivariable" optimisation algorithm implemented in Matlab. Because this paper deals only with the $\text{SNR}_{\text{NLI}}$, this optimisation is based on the received power and pump powers as the figures of merit (and not the total $\text{SNR}$). In this algorithm, the cost function considered is $\sum_{p} P_p$, such that the total power of the pumps is minimised. The optimisation is done over a single span. A nonlinear constraint is also considered such that the received per-channel launch power is above a given threshold. Over the E- and S-band we place 15 pumps spaced from 1~THz apart and let the algorithm find the best power allocation. The highest-wavelength pump was chosen to be 2~THz away from the lowest-wavelength channel.

Ideal amplification is assumed such that the received power is equal to the transmitted power. For FW-RA, pumps are optimised such that at least a quarter of the launch power is recovered at the receiver, for BW-RA and FW+BW-RA, pumps are optimised such that at least half of the launch power is recovered at the receiver. The remaining launch power can be recovered, for instance, with lumped amplification. An example of fully recovered launch power using RA can be found in~\cite{bugliaOFC}. For all scenarios, the pumps' allocation with non-zero power found by the described algorithm is shown in Table~\ref{tab:pumps}. 

For the three scenarios, the per-channel power profile along the distance, i.e., the solution of Eq.~\eqref{eq:diff_Raman}, are shown in Fig.~\ref{fig:profile} for (a) FW-RA, (b) BW-RA and (c) FW+BW-RA cases. Note that, for the FW-RA lower per-channel launch power is chosen (-4~dBm) to limit the per-channel-power peak along the distance to less than 4~dBm as shown in in~Fig.~\ref{fig:profile}~(a); for this case, such high power may be impractical in currently deployed systems, but still, we keep this scenario as a stress-test of the proposed NLI model.

\begin{figure*}[h]
\vspace{-0.5cm}
 \centering
    \input{Figures/fitting.tex}
\vspace{-0.5cm}
\caption{Signal power evolution along the fibre distance obtained using the numerical solution of the Raman differential equations in Eq.~\eqref{eq:diff_Raman} and the semi-analytical solution shown in Eq.~\eqref{eq:Raman_taylor} for (a) FW-RA at 1513.25~nm, FW+BW-RA at 1591.1~nm and (b) BW-RA at 1512.49~nm. In all cases, the results are shown for the worst-performing channel in terms of accuracy between Eq.~\eqref{eq:diff_Raman} and Eq.~\eqref{eq:Raman_taylor}.}\label{fig:fitting}
\end{figure*}
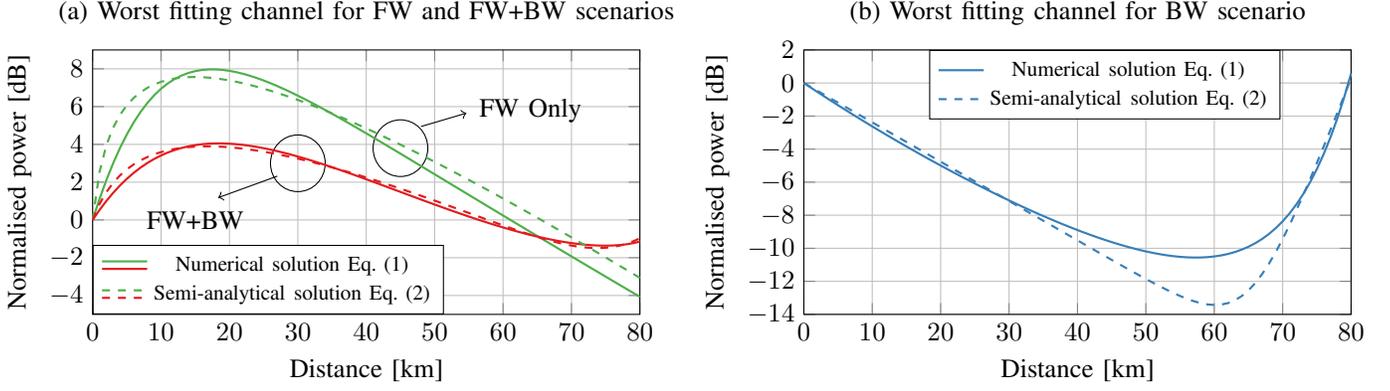

Our goal is now to reproduce the profiles shown in Fig.~\ref{fig:profile}, obtained from Eq.~\eqref{eq:diff_Raman} using the semi-analytical solution shown in Eq.~\eqref{eq:Raman_taylor} after the fitting optimisation routine described in Sec~\ref{sec:signal_profile}. For better visualisation, Fig.~\ref{fig:fitting} shows the results for the worst-performing channel in terms of accuracy between Eq.~\eqref{eq:diff_Raman} and Eq.~\eqref{eq:Raman_taylor} for (a) FW-RA, (b) BW-RA and (c) FW+BW-RA. 

Note that, for the NLI estimation, the effect of the normalised signal profile for each channel is taken into account as an integration over the fibre length (see Eq.~\eqref{eq:link_function_integral}); this means that the inaccuracies shown in Fig.~\ref{fig:fitting} have a negligible impact on the accuracy of the NLI estimation, which is validated in the next section. This is because,
for the FW-RA case (Fig.~\ref{fig:fitting}(a) green), the overestimation of power shown in the first 10~km of fibre is compensated by an underestimation in the remaining kilometres, while for the BW-RA case (Fig.~\ref{fig:fitting}(b)) the inaccuracies occur only for reduced-power levels which do not contribute significantly to the result of the integral in Eq.~\eqref{eq:link_function_integral}. Thus, Fig.~\ref{fig:fitting} shows that the proposed fitting strategy enables reproducing Eq.~\eqref{eq:diff_Raman} by using Eq.~\eqref{eq:Raman_taylor} and accurately capturing the most impactful contributions to the integral in Eq.~\eqref{eq:link_function_integral}.

\section{The closed-form expression for the NLI estimation}
\label{derivation}

This section describes the closed-form expression used to estimate the NLI in the presence of RA. The integral expressions used as a baseline to derive the closed-form expression are presented in Sec.~\ref{sec:integral_expression}. As we will see, these expressions depend on the normalised signal power profile evolution $\rho(z,f_i)$, which were derived in Sec.~\ref{sec:signal_profile}. Thus, Eq.~\eqref{eq:Raman_taylor} is of fundamental importance to derive the closed-form expressions shown in Sec.~\ref{sec:closedform_expression}. This section ends with the application of the closed-form expression in a transmission system and the verification of its accuracy in Sec.~\ref{sec:results_nli}. 

Let $i$ indicate the channel index, the nonlinear signal-to-noise ratio, $\text{SNR}_{\text{NLI},i}$ is given by 
\begin{equation}
   \text{SNR}_{\text{NLI},i} = \frac{P_i}{\eta_n(f_i)P_i^3}, 
\label{eq:snr}
\end{equation}
where $P_i$ is the launch power of the COI and $\eta_n(f_i)$ is the nonlinear coefficient obtained at the end of the $n$\ts{th} span. Note that, as this paper aims to derive a model for NLI noise in Raman-amplified links, we do not extend our analysis to include the amplified spontaneous emission noise, which is left for future work. 

\subsection{The Integral Expressions}
\label{sec:integral_expression}

The integral expressions used to derive the proposed closed-form expressions are as follows. The nonlinear coefficient $\eta_{n}(f_i)$ in Eq.~\eqref{eq:snr}, can be rewritten as~\cite{danielclosed}
\begin{equation}
\begin{split}
&\eta_{n}(f_i) \approx \sum_{j=1}^{n} \left[ \frac{P_{i,j}}{P_i} \right]^2 \cdot [\eta_{\text{SPM}_j}(f_i)n^{\epsilon} + \eta_{\text{XPM}_j}(f_i)],
\label{eq:XPM_SPM_gauss}
\end{split}
\end{equation}
where $\eta_{\text{SPM}_j}(f_i)$ is the self-phase modulation (SPM) contribution and $\eta_{\text{XPM}_j}(f_i)$ is the total cross-phase modulation (XPM) contribution to the NLI both generated in the $j$\ts{th} span. $P_{i,j}$ is the power of channel $i$ launched into the $j$\ts{th} span, $\epsilon$ is the coherent factor~\cite[Eq.~22]{gnmodel}. In Eq.~\eqref{eq:XPM_SPM_gauss}, the four-wave mixing (FWM) contributions to the NLI are neglected, the SPM is assumed to accumulate coherently along the fibre spans, while the XPM is assumed to accumulate incoherently - the accuracy of these assumptions was validated in~\cite{danielclosed}. For notation convenience, the $j$ dependence of the SPM and XPM contribution is suppressed throughout this paper.

The XPM contribution ($\eta_{\text{XPM}}(f_i)$) in Eq.~\eqref{eq:XPM_SPM_gauss} is obtained by summing over all COI-interfering pairs present in the transmitted signal, i.e.,
\begin{equation}
\eta_{\text{XPM}}(f_i)  = \sum_{k = 1, k \neq i}^{N_{ch}} \eta_{\text{XPM}}^{(k)}(f_i),
\label{eq:XPM_gauss}
\end{equation}
where $N_{\text{ch}}$ is the number of WDM channels and $\eta_{\text{XPM}}^{(k)}(f_i)$ is the XPM contribution of a single interfering channel $k$ on channel $i$. 

The XPM and SPM contributions of a single interfering channel are given respectively by~\cite[Eq. 8,9]{danielclosed}
\begin{equation}
\begin{splitfit}
&\eta_{\text{XPM}}^{(k)}(f_i) =\frac{32}{27}\frac{\gamma^2}{B_k^2} \left( \frac{P_k}{P_i} \right)^2 \times \\
& \times \int_{\frac{-B_i}{2}}^{\frac{B_i}{2}} df_1 \int_{\frac{-B_k}{2}}^{\frac{B_k}{2}} df_2 \ \Pi \left(\frac{f_1 + f_2}{B_k} \right)\left| \mu (f_1+f_i, f_2 + f_k, f_i) \right|^2,
\label{eq:XPM_GN_integral}
\end{splitfit}
\end{equation}
and
\begin{equation}
\begin{split}
\eta_{\text{SPM}}(f_i) = \frac{1}{2}\eta_{\text{XPM}}^{(i)}(f_i),
\label{eq:SPM_GN_integral}
\end{split}
\end{equation}
where $\gamma$ is the nonlinear parameter, $\Pi(x)$ denotes the rectangular function and $B_k$ is the bandwidth of the channel $k$. $\mu (f_1, f_2, f_i)$ is the so-called link function or FWM efficiency~\cite{gnmodel}, which is given by~\cite[Eq.~4]{isrsgnmodel}   
\begin{equation}
\begin{split}
&\mu\left(f_1,f_2,f_i\right)=\\ 
&=  \left| \int_0^L d\zeta \ \sqrt{\frac{\rho(\zeta,f_1) \rho(\zeta,f_2) \rho(\zeta,f_1 + f_2 - f_i)}{\rho(\zeta,f_i)}} e^{j\phi\left(f_1,f_2,f_i\right)\zeta}\right|^2
\label{eq:link_function_integral}
\end{split}
\end{equation}
where $\phi=-4\pi^2\left(f_1-f_i\right)\left(f_2-f_i\right)\left[\beta_2+\pi\beta_3(f_1+f_2\right)]$, and $\rho(z,f_i)$ is the normalized signal power profile (see Sec.~\ref{sec:signal_profile}). 
$\beta_2$ is the group velocity dispersion (GVD) parameter, $\beta_3$ is the
linear slope of the GVD parameter. 

\subsection{The derivation of the closed-form expression}
\label{sec:closedform_expression}
This section is devoted to the calculation of $\eta_n(f_i)$ in closed-form, which is then used to calculate $\text{SNR}_{\text{NLI},i}$ in Eq.~\eqref{eq:snr}. The new closed-form expression supporting RA is presented. The formula is obtained by using the semi-analytical solution of the power evolution, obtained in Eq.~\eqref{eq:Raman_taylor} to derive a closed-form expression of the NLI. 

The first step is to derive a closed-form expression of the link function shown in Eq.~\eqref{eq:link_function_integral}. Let
\begin{equation*}
\begin{aligned}[c]
&T_{f,i} = -\frac{P_fC_{f,i}(f_i-\hat{f})}{\alpha_{f,i}}\\
&T_{b,i} = -\frac{P_bC_{b,i}(f_i-\hat{f})}{\alpha_{b,i}}\\
\end{aligned}
\qquad
\begin{aligned}[c]
T_i &= 1 + T_{f,i} - T_{b,i}e^{-\alpha_{b,i}L}\\
\alpha_{l,i} &= \alpha_i + l_1 \alpha_{f,i} - l_2 \alpha_{b,i}\\
\kappa_{f,i} &= e^{-(\alpha_i + l_1\alpha_{f,i})L}\\
\kappa_{b,i} &= e^{-l_2\alpha_{b,i} L}\\
\end{aligned}
\end{equation*}

The link function is approximated in closed-form as
\begin{multline}
\mu\left(f_1 + f_i,f_2 + f_i,f_i\right)\approx
\\\approx \sum_{\substack{0 \leq l_1 + l_2 \leq 1 \\ 0 \leq l_1^\prime + l_2^\prime \leq 1}} \Upsilon_i \Upsilon_i^\prime  \left[  \frac{(\kappa_{f,i} \kappa_{f,i}^\prime + \kappa_{b,i} \kappa_{b,i}^\prime)( \alpha_{l,i} \alpha_{l,i}^\prime + \phi^2)}{(\alpha_{l,i}^2 + \phi^2)(\alpha_{l,i}^{\prime 2} + \phi^2)} \right.-\\
- \frac{(\kappa_{f,i} \kappa_{b,i}^\prime +  \kappa_{b,i} \kappa_{f,i}^\prime)( \alpha_{l,i} \alpha_{l,i}^\prime + \phi^2) }{(\alpha_{l,i}^2 + \phi^2)(\alpha_{l,i}^{\prime 2} + \phi^2)}\cos(\phi L) +\\
\left. + \frac{(\kappa_{f,i} \kappa_{b,i}^\prime -  \kappa_{b,i} \kappa_{f,i}^\prime)( \alpha_{l,i} - \alpha_{l,i}^\prime) \phi }{(\alpha_{l,i}^2 + \phi^2)(\alpha_{l,i}^{\prime 2} + \phi^2)}\sin(\phi L)
\right],
\label{eq:link_function_closed}
\end{multline}
where $\Upsilon_i$ is given by
\begin{equation}
\Upsilon_i
= T_i \left(\frac{-T_{f,i} }{T_i} \right)^{l_1} \left(\frac{T_{b,i} }{T_i} \right)^{l_2}.
\label{eq:Upsilon_link_function}
\end{equation}
The proof of Eq.~\eqref{eq:link_function_closed} is given in Appendix~\ref{appB:link_function}. The coefficient $\Upsilon_i^\prime$ is respectively the same as the one in Eq.~\eqref{eq:Upsilon_link_function} with the indices $l_1$ and $l_2$ replaced by $l_1^\prime$ and $l_2^\prime$. The same is valid for the variables $\alpha_{l,i}^\prime$, $\kappa_{f,i}^\prime$ and $\kappa_{b,i}^\prime$.

We now present a closed-form expression for the XPM and SPM NLI contributions shown in Eqs.~\eqref{eq:XPM_GN_integral} and \eqref{eq:SPM_GN_integral}, respectively. Using Eq.~\eqref{eq:link_function_closed} as an analytical solution of the link function, a closed-form expression for the XPM and SPM are given respectively by
\hspace{-1.5cm}
\begin{equation}\label{eq:XPM_closed}
\begin{splitfit}
&\eta_{\text{XPM}}^{(k)}(f_i) =  \frac{32}{27} \frac{\gamma^2}{B_k} \left(\frac{P_k}{P_i}\right)^2 \sum_{\substack{0 \leq l_1 + l_2 \leq 1 \\ 0 \leq l_1^\prime + l_2^\prime \leq 1}} \Upsilon_k \Upsilon_k^\prime\frac{1}{\phi_{i,k} (\alpha_{l,k} + \alpha_{l,k}^\prime)}\times\\
&\times \Bigggl\{2(\kappa_{f,k} \kappa_{f,k}^\prime + \kappa_{b,k} \kappa_{b,k}^\prime) \left[\atan\!\left(\frac{\phi_{i,k}B_i}{2\alpha_{l,k}}\right)	+\atan\!\left(\frac{\phi_{i,k}B_i}{2\alpha_{l,k}^\prime}\right)\right] +\\
&\qquad+~ \pi \bigggl[ - (\kappa_{f,k} \kappa_{b,k}^\prime +  \kappa_{b,k} \kappa_{f,k}^\prime) \left(  \sign\!\left(\frac{\alpha_{l,k}}{\phi_{i,k}} \right)  e^{-|\alpha_{l,k}L|} \right. +\\
&\qquad+~  \left. \sign\!\left(\frac{\alpha_{l,k}^\prime}{\phi_{i,k}} \right)  e^{-|\alpha_{l,k}^\prime L|}\right) + (\kappa_{f,k} \kappa_{b,k}^\prime - \kappa_{b,k} \kappa_{f,k}^\prime) \times\\
&\qquad \times~
	\left(  \sign\!\left(-\phi_{i,k} \right)  e^{-|\alpha_{l,k}L|} +  \sign\!\left(\phi_{i,k} \right)  e^{-|\alpha_{l,k}^\prime L|}\right) \bigggr] \Bigggr\}
\end{splitfit}
\end{equation}
and
\begin{equation}\label{eq:SPM_closed}
\begin{splitfit}
&\eta_{\text{SPM}}(f_i) = 
 \frac{16}{27} \frac{\gamma^2}{B^2_i}\sum_{\substack{0 \leq l_1 + l_2 \leq 1 \\ 0 \leq l_1^\prime + l_2^\prime \leq 1}} \Upsilon_i \Upsilon_i^\prime \frac{\pi}{\phi_{i}(\alpha_{l,i} + \alpha_{l,i}^\prime)} \times\\
 &\times \Bigggl\{ 2(\kappa_{f,i} \kappa_{f,i}^\prime + \kappa_{b,i} \kappa_{b,i}^\prime)  \left[\asinh\!{\left(\frac{3 \phi_i B_i^2}{8 \pi\alpha_{l,i}}\right)}  + \asinh\!{\left(\frac{3 \phi_i B_i^2}{8 \pi \alpha_{l,i}^\prime}\right)} \right] + \\
 & + 4 \ln\!\left(\sqrt{\frac{\phi_i L}{2\pi}} B_i  \right) \bigggl[  - (\kappa_{f,i} \kappa_{b,i}^\prime +  \kappa_{b,i} \kappa_{f,i}^\prime)  \left( \sign\!\left(\frac{\alpha_{l,i}}{\phi_{i}} \right)  e^{-|\alpha_{l,i}L|} + \right. \\
 & + \left. \sign\!\left(\frac{\alpha_{l,i}^\prime}{\phi_{i}} \right)  e^{-|\alpha_{l,i}^\prime L|}\right) + (\kappa_{f,i} \kappa_{b,i}^\prime - \kappa_{b,i} \kappa_{f,i}^\prime ) \times \\
&\times \left(  \sign\left( - \phi_i \right)  e^{-|\alpha_{l,i}L|}  \sign\left(\phi_i \right)  e^{-|\alpha_{l,i}^\prime L|}\right) \bigggr]  \Bigggr\},
\end{splitfit}
\end{equation}
where
\begin{align*}
 \phi_i&=-4\pi^2\left(\beta_2+2\pi\beta_3f_i\right),\\
 \phi_{i,k}&=-4\pi^2\left(f_k-f_i\right)\left[\beta_2+\pi\beta_3\left(f_i+f_k\right)\right].
\end{align*} 
The proof of Eqs.~\eqref{eq:XPM_closed} and \eqref{eq:SPM_closed} are given respectively in Appendix~\ref{appC:XPM}~and~\ref{appD:SPM}.

Finally, the $\SNR_{\NLI,i}$ can be calculated analytically by inserting Eqs.~\eqref{eq:XPM_SPM_gauss},~\eqref{eq:XPM_gauss},~\eqref{eq:XPM_closed} and \eqref{eq:SPM_closed} in Eq.~\eqref{eq:snr}. The final expression accounts for wavelength-dependent fibre parameters and different launch power per channel. Additionally, the formula is also valid for links made of different span setups - in that case, all the fibre parameters and per-channel launch power depend not only on the channel~$i$ but also on the span~$j$.

\subsection{Results for the nonlinear interference estimation}
\label{sec:results_nli}

This section shows the validation of Eq.~\eqref{eq:SPM_closed} and Eq.~\eqref{eq:XPM_closed}. To that end, we consider the transmission system described in Sec.~\ref{sec:signal_profile}, i.e., a distributed RA link consisting of a WDM transmission with $N_{\text{ch}}$=131 channels spaced by 100~GHz and centred at 1550~nm. Each channel was modulated at the symbol rate of 96~GBd, 
resulting in a total bandwidth of 13~THz~(105~nm). Gaussian symbols are considered in the transmission. The span length is 80~km and an ITU-T G652.D fibre is considered with Raman gain spectrum and attenuation shown in Fig.~\ref{fig:raman_loss}. Nonlinear coefficient and dispersion parameters are $\gamma = 1.16\text{ W}^{-1} \text{km}^{-1}$,  $D = 16.5~\text{ps }\text{nm}^{-1}\text{km}^{-1}$, $S = 0.09~\text{ps }\text{nm}^{-2}\text{km}^{-1}$, respectively. A spectrally uniform
launch power profile, where each channel carries -4~dBm, 0~dBm and -2~dBm is considered respectively for FW-RA,
BW-RA and FW+BW-RA (see Sec.\ref{sec:results_signal_profile}). The power profiles along the fibre distance are shown in Fig.~\ref{fig:profile} and the pumps' allocation used for each one of the scenarios is shown in Table~\ref{tab:pumps}. Results are obtained for single-span, 3-span and 10-span transmissions. The amplifiers are assumed to be ideal, such that the launch power profile is the same at the beginning of each span and equal to the transmitted power.

\begin{figure}[t!]
\input{Figures/SNR.tex}
\caption{Nonlinear performance after 1~x~80~km, 3~x~80~km and 10~x~80~km transmission for (a) FW-RA, (b) BW-RA and (c) FW+BW-RA.}
\label{fig:SNR}
\end{figure}
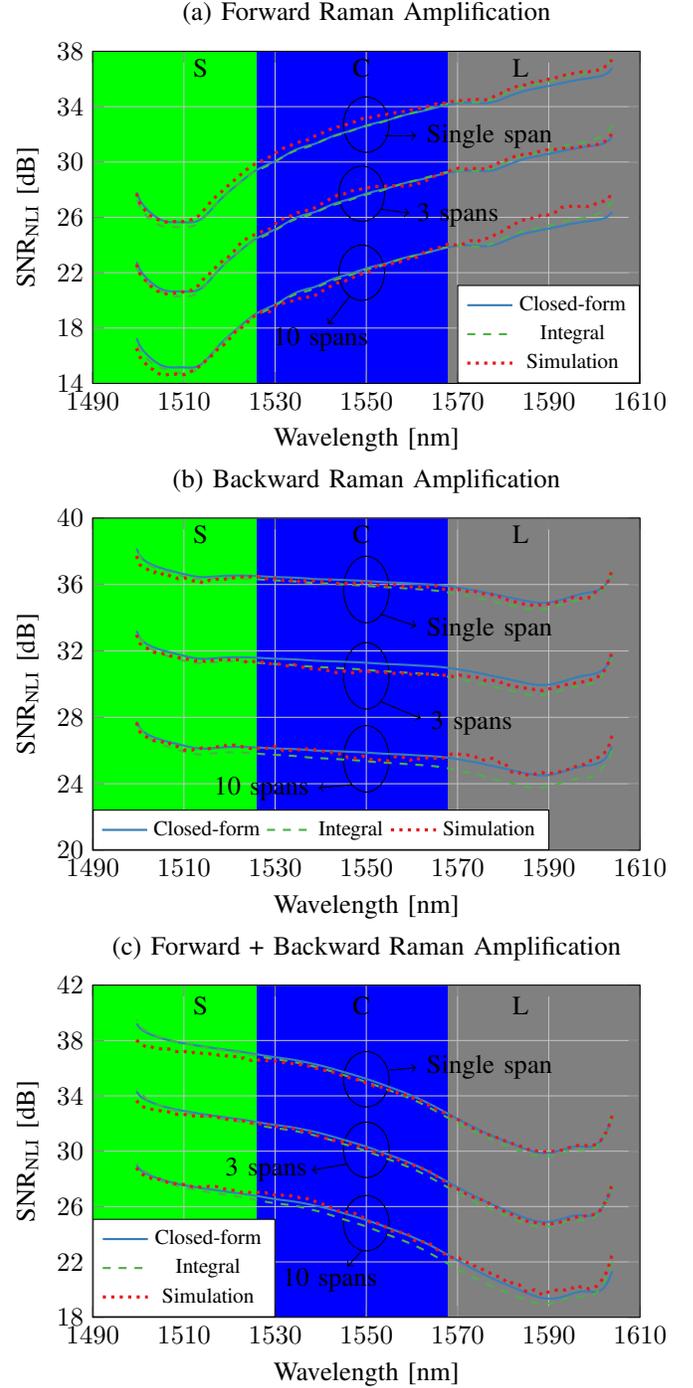

The $\text{SNR}_{\text{NLI}}$ as a function of wavelength is shown in Fig.~\ref{fig:SNR} for (a) FW-RA, (b) BW-RA and (c) FW+BW-RA for the cases of a single span, 3-span and 10-span transmissions. To verify the accuracy of the closed-form expression shown in Eqs.~\eqref{eq:XPM_closed} and~\eqref{eq:SPM_closed}, the $\text{SNR}_{\text{NLI}}$ is also computed using the integral ISRS GN model~\cite{isrsgnmodel} and SSFM simulations. For the former, the results are obtained by inserting the power profiles shown in Fig.~\ref{fig:profile} in~\cite[Eq.~4]{isrsgnmodel}. For the latter, the same power profiles from Fig.~\ref{fig:profile} are used and interpolated along the fibre distance for each step in the SSFM simulation. To ensure accurate simulation results, adaptive step sizes with local-error method~\cite{sinkinOptimizationSplitstepFourier2003} was used, where goal local error $\delta_G=10^{-10}$ and sequence of $2^{17}$ Gaussian symbols per channel were considered. Note that, for all the results, the XPM generated by the pumps is neglected; as shown in~\cite{Iqbal:20}, this is a valid assumption when the WDM spectra are sufficiently far from the pumps - in our case, as described in Sec.~\ref{sec:results_nli}, the highest-wavelength pump was chosen to be 2 THz away from the lowest-wavelength channel, such that these effects could be neglected. Despite that, the aforementioned effects can be included in this model by considering the pumps as additional interfering channels.

Fig.~\ref{fig:SNR} shows the $\text{SNR}_{\text{NLI}}$ for (a) FW-RA, (b) BW-RA and (c) FW+BW-RA. It is interesting to note the correlation of the $\text{SNR}_{\text{NLI}}$ profile with the power profiles shown in Fig.~\ref{fig:profile}. Indeed, for the FW-RA case, shown in Fig.~\ref{fig:SNR}(a), the high-power levels in short wavelengths (see Fig.~\ref{fig:profile}(a)), reduce the $\text{SNR}_{\text{NLI}}$, degrading the performance of those channels; on the other hand, the performance of long-wavelength channels is higher, due to their reduced power levels, yielding to a tilt in the $\text{SNR}_{\text{NLI}}$ profile. For the BW-RA case, shown in Fig.~\ref{fig:SNR}(b) the interaction between fibre attenuation, dispersion and power profile (see Fig.~\ref{fig:profile}(b)) yields a relatively flat $\text{SNR}_{\text{NLI}}$ profile; however, a smooth tilt can still be observed, which also correlate with the power profile shown in Fig.~\ref{fig:profile}(b) as high-power levels are observed for the longer wavelengths. Note that, in general, BW-RA performs better in terms of $\text{SNR}_{\text{NLI}}$ when compared with FW-RA case because of the reduced per-channel power evolution along the fibre. Finally, for the FW+BW-RA case shown in Fig.~\ref{fig:SNR}(c), the increased power levels at higher wavelengths (see Fig.~\ref{fig:profile}(c)), which is a result of the FW pumping, degrades the performance of those channels when compared to the lower wavelengths channels, where BW pumping dominates. This results in higher $\text{SNR}_{\text{NLI}}$ values for channels located in the S-band when compared to those in the L-band, also yielding a tilt in the $\text{SNR}_{\text{NLI}}$ profile.

In terms of accuracy, for a single-span FW-RA transmission, maximum per-channel errors of 0.81~dB and 0.64~dB were found between the closed-form expression and the integral ISRS GN model, and between the closed-form expression and the SSFM simulation, respectively. For the transmission over 3 spans, these errors are respectively 0.78~dB and 0.61~dB. For the transmission over 10 spans, these errors are respectively 0.74~dB and 1.47~dB.
The same analyses for the BW-RA transmission over a single span yield errors of 0.47~dB and 0.54~dB respectively, while for the transmission over 3 spans, these errors are both equal to 0.67~dB. Over 10 spans, these errors are respectively 0.80~dB and 0.68~dB.
Finally, the same analysis for the FW+BW-RA over a single span yield errors of 0.31~dB and 1.18~dB respectively, for the transmission over 3 spans, these errors are respectively 0.41~dB and 0.65~dB, and for transmission over 10 spans, these errors are 0.68~dB and 1.23~dB.

\section{Conclusions}

In this work, we presented a closed-form formula of the Gaussian noise (GN) model suitable for ultra-wideband (UWB) transmission systems and tested over a distributed Raman amplification setup. This formula is the first to account for any setup of Raman amplification technologies together with the inter-channel stimulated Raman scattering (ISRS) effect. The formula is shown to support forward (FW) and backward (BW) pumping schemes and accurately predict the nonlinear interference (NLI) for an arbitrary number of pumps and wavelength-dependent fibre parameters and launch power profiles. A fundamental step to deriving this closed formula was to derive a semi-analytical solution to correctly reproduce the signal power profile evolution along the fibre distance in the presence of Raman amplification and ISRS effect. 

The formula was applied to 13~THz optical bandwidth corresponding to transmission over the S-, C-, and L- bands. In terms of accuracy, among all of the scenarios tested in this work, the formula showed maximum errors of 0.81~dB and 1.47~dB when compared to the integral model and slit-step Fourier method (SSFM) simulations.
Additionally, the formula is capable of estimating the NLI in only a few seconds, where the majority of the
computational time was required to numerically solve the differential Raman equations. Because of the speed of computation, the formula is suitable for real-time estimation of the NLI and can be applied as an enabling tool for future intelligent and dynamic optical fibre networks.

\section*{Data Availability Statement}
The data that support the figures in this paper are available from the UCL Research Data Repository (DOI:10.5522/04/21696401), hosted by FigShare.

\appendices
\section{Derivation of the Analytical Solution of the Normalized Signal Power Profile.}
\label{appA:triangular_approximation}

This section shows the derivation of Eq.~\eqref{eq:Raman_taylor}. We stress that most of the assumptions made in this section are not exact, however, this is not an issue as this equation is used as a semi-analytical solution of the Raman equations and the coefficients will be fitted and optimised. 

We start with Eq.~\eqref{eq:diff_Raman}. The derivation is analogous to~\cite{zirngibl1998analytical}. We start by considering a constant attenuation $\alpha$ for all the channels and neglecting the energy that is lost whenever a high-frequency photon
is transformed into a low-frequency photon, i.e., $\frac{f_k}{f_i} \approx 1$ and $\frac{f_p}{f_i} \approx 1$. Also, we assume the triangular approximation of the Raman spectrum, i.e., $g_r(\Delta f) \approx C_r \Delta f$, where $C_r$ is the slope of the linear regression (normalized by the effective core area $A_\text{eff}$) and $\Delta f$ is the frequency separation between the channels and between the channels and the pumps. Under these assumptions, Eq.~\eqref{eq:diff_Raman} can be written as

\begin{equation}
\begin{aligned}
  &\frac{\partial P_i}{\partial z} = \\ &=\sum_{k=1}^{N_{ch}} C_r (f_k - f_i) P_k P_i + \sum_{p=1}^{N_p} C_r (f_p - f_i) P_k P_i - \alpha P_i = \\
  &=  C_r P_i  \left (\sum_{k=1}^{N_{ch}} (f_k - f_i) P_k + \sum_{p=1}^{N_p}(f_p - f_i) P_k \right) - \alpha P_i.
\end{aligned}
\label{eq:diff_Raman_app}
\end{equation}
We now write the coupled differential equations into one equation, by replacing the $N_{ch}$ signals and $N_p$ pumps by a signal and pump density spectrum. Also, we replace the
summation by an integration over the entire frequency spectrum
of the signal and the pumps. Thus, Eq.~\eqref{eq:diff_Raman_app} can be written as

\begin{equation}
\begin{aligned}
 &\frac{d P(z,f)}{d z} = \\
  & =C_r P(z,f)  \left (\int_{f_{ch,min}}^{f_{ch,max}} (\Lambda_{ch} - f) P(z,\Lambda_{ch}) \,d\Lambda_{ch} \right.+\\
  & \left. + \int_{f_{p,min}}^{f_{p,max}}(\Lambda_p - f) P(z,\Lambda_p) \, d\Lambda_p \right) - \alpha P(z,f),
\end{aligned}
\label{eq:diff_Raman_app_int}
\end{equation}
where $f_{ch,max}$, $f_{ch,min}$, $f_{p,max}$ and $f_{p,min}$ are respectively the maximum and minimum frequencies of the channels, and the pumps. Dividing both sides of Eq.~\eqref{eq:diff_Raman_app_int} by $P(z,f)$ and taking the derivative with respect to the frequency $f$, we have
\begin{equation}
\begin{aligned}
&\frac{d}{d f} \left ( \frac{d P(z,f)/d z}{P(z,f)} \right )=- C_r \times \\
  &  \times \left (\underbrace{\int_{f_{ch,min}}^{f_{ch,max}} P(z,\Lambda_{ch}) \,d\Lambda_{ch}}_\text{$P_{total, ch}$} \right.
  & \left. + \underbrace{\int_{f_{p,min}}^{f_{p,max}} P(z,\Lambda_p) \, d\Lambda_p}_\text{$P_{total, p}$} \right)
\end{aligned}
\label{eq:diff_Raman_app_int2}
\end{equation}
Note that, the integrals represent the total launch power ($P(z)$), i.e., a sum of the channel ($P_{total, ch}$), the forward pump ($P_{total, fw}$) and backward pump ($P_{total, bw}$) launch powers. Moreover, $P_{total, ch}$ and $P_{total, fw}$ must decay with $e^{-\alpha z}$, while $P_{total, bw}$ decays with $e^{-\alpha(L-z)}$. Thus, Eq.~\eqref{eq:diff_Raman_app_int2} can be written as
\begin{equation}
\begin{aligned}
&\frac{d}{d f} \left ( \frac{d P(z,f)/d z}{P(z,f)} \right )= -C_rP(z) =\\
  & = - C_r \, (P_{total, ch}e^{-\alpha z} + P_{total, fw}e^{-\alpha z} + P_{total, bw}e^{-\alpha (L-z)}).
\end{aligned}
\label{eq:diff_Raman_app_int3}
\end{equation}
In order to apply this equation in more general scenarios and overcome the assumptions done so far, we define separate wavelength-dependent attenuation to model channels together with FW pumps ($\alpha_{f,i}$) and together with BW pumps ($\alpha_{b,i}$). These parameters can be interpreted as modelling respectively how fast the channel gain/loss due to the FW-RA and BW-RA together with ISRS extinguishes along the fibre. Similarly, we define separate wavelength-dependent $C_r$, i.e., $C_{f,i}$ and $C_{b,i}$, respectively for channels together with FW pumps and BW pumps. These two parameters model respectively the gain/loss due to FW-RA and BW-RA together with the ISRS effect. 
Note that, a more rigorous approach would be to define 3 wavelength-dependent $C_r$ and $\alpha$ for each one of the terms on the right-hand side of Eq~\eqref{eq:diff_Raman_app_int3} (as they occupy different locations of the frequency spectrum); however, as ISRS + FW pumping have effectively the same effect (a power transfer from pumps/channels to the COI) they can be modelled joint, resulting in a total of 2 wavelength-dependent $C_r$ and $\alpha$. 
Thus, by letting $P_f = P_{total, ch} + P_{total, fw}$ and $P_b = P_{total, bw}$, Eq.~\eqref{eq:diff_Raman_app_int3} is rewritten as
\begin{multline}
  \frac{d}{d f} \left ( \frac{d P(z,f)/d z}{P(z,f)} \right ) =\\
  = - (C_{f,i} P_f e^{-\alpha_{f,i} z} + C_{b,i}P_be^{-\alpha_{b,i} (L-z)}).
\label{eq:diff_Raman_app_int4}
\end{multline}
Now, we integrate with respect to $z$ and $f$. For the integration in $f$, note that, the WDM spectra and the pumps occupy different parts of the frequency spectrum. This fact massively complicates the derivation of the closed-form expression. To tackle this, without loss of generality, let us consider a new central frequency (common to the WDM spectra and the pumps) as the average frequencies of the pumps, which we denote by $\hat{f}$ (other choices are also possible). Thus, integrating over $z$ and $f$ yields
\begin{equation}
\begin{aligned}
  &P(z,f)=
  &e^{- [C_{f,i} P_{f} L_\text{eff}(f-\hat{f}) + C_{b,i} P_b \tilde{L}_\text{eff}(f-\hat{f})]+A(z) + B(f)},
\end{aligned}
\label{eq:diff_Raman_app_int5}
\end{equation}
where $L_\text{eff} = \frac{1-e^{-\alpha_{f,i} z}}{\alpha_{f,i}}$ and $\tilde{L}_\text{eff} = \frac{e^{-\alpha_{b,i}(L-z)}-e^{-\alpha_{b,i} L}}{\alpha_{b,i}}$, and $A(z)$, $B(f)$ arbitrary functions which their values determined by requiring that $P(z=0,f) = P(0,f)$, which immediately implies that $e^{B(f)}=P(0,f)$, and, by requiring $\int P(z,f)\,df = P(z)$, the value of $e^{A(z)}$ is obtained. 
The evolution of the power $P(z)$ is assumed to be modelled as $P(z) = P_{total}e^{-\alpha_i z}$. This is not strictly true as a power gain at the end of the fibre is expected because of the presence of BW pumps. However, in addition to facilitating the derivation of the closed-form expression, this assumption is also overcome by the wavelength-dependent fitting parameters ${\alpha_{b,i}}$ and $C_{b,i}$, which are responsible for giving gains at the end of the fibre. 
Thus, Eq.~\eqref{eq:diff_Raman_app_int5} is written as
\begin{multline}
\rho(z,f) = \frac{P(z,f)}{P(0,f)} =\\
= \frac{P_{total}e^{-\alpha_i z}e^{-(C_{f,i}P_{f}L_\text{eff} + C_{b,i} P_{b}\tilde{L}_\text{eff})(f-\hat{f})}}{\int G_{\text{Tx}}(\nu)e^{-(C_{f,i} P_{f}L_\text{eff} + C_{b,i} P_{b}\tilde{L}_\text{eff})\nu} d\nu},
\label{appB:eq:diff_Raman_app_int6}
\end{multline}
where $G_{Tx}(f)$ is the input signal spectra including the WDM channels and the pumps and $P_{total}$ is the sum of its launch power. Moreover, the coefficient $\alpha$ is also considered a wavelength-dependent loss $\alpha_i$. Let
$x_i = C_{f,i}P_{f}L_\text{eff} + C_{b,i}P_{b}\tilde{L}_\text{eff}$. 
By assuming that the input power $G_{Tx}(f)$ is uniformly distributed over the optical bandwidth $B$ with power $P_{total}$ we can write, 
\begin{equation}
\begin{aligned}
\int G_{\text{Tx}}(\nu)e^{-x_i\nu} d\nu = \frac{2P_{total}\sinh{\big(\frac{x_iB}{2}}\big)}{x_iB}.
\end{aligned}
\label{appB:eq:diff_Raman_app_int7}
\end{equation}
Replacing Eq.\eqref{appB:eq:diff_Raman_app_int7} in Eq.\eqref{appB:eq:diff_Raman_app_int6} leads to
\begin{equation}
\begin{aligned}
&\rho(z,f) =e^{-\alpha_i z} \frac{x_iB e^{-x_i(f-\hat{f})}}{2\sinh{\big(\frac{xB}{2}\big)}}. 
\end{aligned}
\label{appB:eq:diff_Raman_app_int8}
\end{equation}
Finally, by expanding Eq.~\eqref{appB:eq:diff_Raman_app_int8} using a 1\ts{st} order Taylor approximation around the point $x_i = 0$, yields
\begin{equation}
\begin{aligned}
&\rho(z,f) = e^{-\alpha_i z}[1 - x(f - \hat{f})], 
\end{aligned}
\label{appB:eq:diff_Raman_app_int9}
\end{equation}
and setting $f=f_i$, Eq.~\eqref{eq:Raman_taylor} is obtained concluding the proof.

In the derivation of Eq.~\eqref{appB:eq:diff_Raman_app_int9}, 5 wavelength-dependent coefficients are introduced, namely, $\alpha_i$, $\alpha_{f,i}$, $\alpha_{b,i}$, $C_{f,i}$ and $C_{b,i}$. These coefficients are chosen to overcome the restrictive assumptions used to derive Eq.~\eqref{appB:eq:diff_Raman_app_int9} and to enable a simplified derivation of the closed-form expression in the following Appendices. The restrictive assumptions overcome by the aforementioned fitting parameters are: the constant attenuation $\alpha$ for all the channels; the energy that is lost whenever a high-frequency photon
is transformed into a low-frequency photon; the triangular approximation of the Raman spectrum; the different spectrum locations of FW pumps, BW pumps and channels through a new central frequency of the spectrum $\hat{f}$ and the joint consideration of the FW amplification and ISRS effect; the power evolution along the distance $P(z) = P_{total}e^{-\alpha_i z}$; the spectrally uniform input power profile $G_{Tx}(f)$; and the 1\ts{st} order Taylor approximation.

\section{Derivation of the link function.}
\label{appB:link_function}

This section shows the derivation of Eq.~\eqref{eq:link_function_closed}. Let $x_i(\zeta) = C_{f,i}P_{f}L_\text{eff} + C_{b,i}P_{b}\tilde{L}_\text{eff}$, $\tau_i(\zeta) =1 - x_i(\zeta)(f_i - \hat{f})$ with $L_\text{eff}(\zeta) = \frac{1-e^{-\alpha_{f,i} \zeta}}{\alpha_{f,i}}$ and  $\tilde{L}_\text{eff}(\zeta) = \frac{e^{-\alpha_{b,i}(L-\zeta)}-e^{-\alpha_{b,i} L}}{\alpha_{b,i}}$. The first step is to insert Eq.~\eqref{appB:eq:diff_Raman_app_int6} in Eq.~\eqref{eq:link_function_integral} and use the approximation in Eq.~\eqref{appB:eq:diff_Raman_app_int7} yielding
\begin{multline}
\mu\left(f_1,f_2,f_i\right) = \\
=\left| \int_0^L d\zeta \
e^{-\alpha_i \zeta} \frac{x_iB e^{-x_i(f_1 + f_2 - f_i -\hat{f})}}{2\sinh{\big(\frac{x_iB}{2}\big)}}
e^{j\phi\left(f_1,f_2,f_i\right)\zeta}\right|^2.
\label{appC:link_function_integralapp_new}
\end{multline}

Now, we consider the link function for the XPM contribution in Eq.~\eqref{eq:XPM_GN_integral}, i.e., $\mu\left(f_1 + f_i,f_2 + f_k,f_i\right)$ (the derivation for the link function for SPM is analogous and one simply needs to replace $f_k = f_i$ and the indices $k=i$). Assuming that the frequency separation between channels $k$ and $i$ ($\Delta f = f_k - f_i$) is much larger than half of the bandwidth of channel $k$ ($|\Delta f| \gg \frac{B_k}{2}$), we can assume that $f_2 + \Delta f \approx \Delta f$. Also, we assume that the signal
power profile is constant over the channel bandwidth (see Appendices in~\cite{danielclosed} for additional details). Then, using the 1~\ts{st} order Taylor approximation shown in Eq.~\eqref{appB:eq:diff_Raman_app_int9}, yields to
\begin{multline}
\mu\left(f_1 + f_i,f_2 + f_k,f_i\right) = \\
= \left| \int_0^L d\zeta \
e^{-\alpha_k \zeta} \tau_k(\zeta)
e^{j\phi\left(f_1 + f_i,f_2 + f_k,f_i\right)\zeta}\right|^2.
\label{appC:link_function_integral1}
\end{multline}
The term $\tau_k(\zeta)$ can be written as
\begin{multline}
\tau_k(\zeta) = 1 - \left[\left(\frac{C_{f,k}P_{f}}{\alpha_{f,k}}\right) \left(1-e^{-\alpha_{f,k}\zeta}\right)  +\right.\\ 
 + \left. \left(\frac{C_{b,k}P_{b}}{\alpha_{b,k}}\right) e^{-\alpha_{b,k} L} \left(e^{\alpha_{b,k} \zeta}-1\right)\right](f_k - \hat{f}).
\label{appC:link_function_integral2}
\end{multline}
Let \quad $T_{f,k} = \frac{-P_fC_{f,k}(f_k-\hat{f})}{\alpha_{f,k}}$,
\quad $T_{b,k} = \frac{-P_bC_{b,k}(f_k-\hat{f})}{\alpha_{b,k}}$,\\
$T_k = 1 + T_{f,k} - T_{b,k}e^{-\alpha_{b,k}L}$. Thus, the term $\tau_k(\zeta)$ is written as 
\begin{equation}
\begin{split}
&\tau_k(\zeta) = T_k\left[1 - \frac{T_{f,k}}{T_k}e^{-\alpha_{f,k} \zeta} + \frac{T_{b,k}}{T_k}e^{-\alpha_{b,k} L}e^{\alpha_{b,k} \zeta}\right].
\label{appC:link_function_integral3}
\end{split}
\end{equation}
Eq.~\eqref{appC:link_function_integral3} can be conveniently rewritten in terms of a summation using identity \eqref{appE:multinomial_theorem}, which will facilitate all the mathematical derivations,
\begin{multline}
\tau_k(\zeta) = T_k\sum_{\substack{0 \leq l_1 + l_2 \leq 1}} \left( \frac{-T_{f,k}}{T_k} \right)^{l_1} \left( \frac{T_{b,k}}{T_k} \right)^{l_2} \times \\ 
\times e^{-(l_1\alpha_{f,k} \zeta + l_2 \alpha_{b,k} L - l_2 \alpha_{b,k} \zeta)}.
\label{appC:link_function_integral4}
\end{multline}
Now, defining 
\begin{equation}
\begin{split}
&\Upsilon_k
= T_k \left(\frac{-T_{f,k} }{T_k} \right)^{l_1} \left(\frac{T_{b,k} }{T_k} \right)^{l_2},
\label{appC:link_function_integral5}
\end{split}
\end{equation}
Eq.~\eqref{appC:link_function_integral4} is written as 
\begin{equation}
\begin{split}
&\tau_k(\zeta) =  \sum_{\substack{0 \leq l_1 + l_2 \leq 1}} \Upsilon_k e^{-(l_1\alpha_{f,k} \zeta + l_2 \alpha_{b,k} L - l_2 \alpha_{b,k} \zeta)}.
\label{appC:link_function_integral6}
\end{split}
\end{equation}
Note that $\Upsilon_k$ is a variable which depends on the indices of the summation. 
Now, inserting Eq.~\eqref{appC:link_function_integral6} in Eq.~\eqref{appC:link_function_integral1}, we obtain 
\begin{equation}\label{appC:link_function_integral7}
\begin{splitfit}
&\mu\left(f_1 + f_i,f_2 + f_k,f_i\right) = \\
& =\left|\sum_{\substack{0 \leq l_1 + l_2 \leq 1}} \Upsilon_k \int_0^L d\zeta \
 e^{-(\alpha_k \zeta + l_1\alpha_{f,k} \zeta + l_2 \alpha_{b,k} L - l_2 \alpha_{b,k} \zeta) + j \phi \zeta }\right|^2,
\end{splitfit}
\end{equation}
and solving the integral in Eq.\eqref{appC:link_function_integral7} yields to 
\begin{multline}
\mu\left(f_1 + f_i,f_2 + f_k,f_i\right) = \\ 
= \left|\sum_{\substack{0 \leq l_1 + l_2 \leq 1}} \Upsilon_k \frac{e^{-(\alpha_k + l_1\alpha_{f,k})L + j \phi L} - e^{-l_2\alpha_{b,k} L}}{-(\alpha_k + l_1 \alpha_{f,k} - l_2 \alpha_{b,k}) + j \phi}\right|^2.
\label{appC:link_function_integral8}
\end{multline}
Now, let define $\alpha_{l,k} = \alpha_k + l_1 \alpha_{f,k} - l_2 \alpha_{b,k}$, $\kappa_{f,k} = e^{-(\alpha_k + l_1\alpha_{f,k})L}$ and $\kappa_{b,k} = e^{-l_2\alpha_{b,k} L}$. Eq.~\eqref{appC:link_function_integral8} can then be written as 
\begin{equation}
\begin{split}
&\mu\left(f_1 + f_i,f_2 + f_k,f_i\right) =  \left|\sum_{\substack{0 \leq l_1 + l_2 \leq 1}} \Upsilon_k \frac{\kappa_{f,k} e^{j \phi L} - \kappa_{b,k}}{- \alpha_{l,k} + j \phi}\right|^2.
\label{appC:link_function_integral9}
\end{split}
\end{equation}
The last step of the derivation is to calculate the modulus of Eq.~\eqref{appC:link_function_integral9}. Using the identity \eqref{appE:modulo_complex_number} we can write Eq.~\eqref{appC:link_function_integral9} as
\begin{equation}
\begin{splitfit}
&\mu\left(f_1 + f_i,f_2 + f_k,f_i\right)= \left ( \sum_{\substack{0 \leq l_1 + l_2 \leq 1}} \Upsilon_k \frac{\kappa_{f,k} e^{j \phi L} - \kappa_{b,k}}{- \alpha_{l,k} + j \phi} \right)  \times\\
& \times \left ( \sum_{\substack{0 \leq l_1^\prime + l_2^\prime \leq 1}} \Upsilon_k^\prime \frac{\kappa_{f,k}^\prime e^{-j \phi L} - \kappa_{b,k}^\prime}{- \alpha_{l,k}^\prime - j \phi}\right).
\label{appC:link_function_integral10}
\end{splitfit}
\end{equation}
Finally, performing the multiplication in Eq.~\eqref{appC:link_function_integral10} together with the identity \eqref{appE:sum_modulo_complex_number} and considering the channel $f_k = f_i$ yields to Eq.~\eqref{eq:link_function_closed}, concluding the proof.

\section{Derivation of the XPM contribution.}
\label{appC:XPM}

This section shows the derivation of Eq.~\eqref{eq:XPM_closed}. We start by approximating the phase mismatch term in Eq.~\eqref{eq:link_function_integral}.  For the XPM contribution, let $\Delta f = f_k - f_i$ be the frequency separation between channels $k$ and $i$ - here the pumps are also included as additional indices $k$. Assuming that frequency
separation is much larger than half of the bandwidth of channel $k$ ($|\Delta f| \gg \frac{B_k}{2}$), we can make the assumption that $f_2 + \Delta f \approx \Delta f$. Also, we assume that the dispersion slope $\beta_3$ is constant over the channel bandwidth. Thus, the phase mismatch term can be approximated as~\cite[Eq. 15]{danielclosed},

\begin{equation}
\begin{split}
&\phi(f_1+f_i,f_2+f_k,f_i)=\\
& = -4\pi^2 f_1 \Delta f \left[\beta_2+\pi\beta_3(f_1+f_2+f_i+f_k\right)]\approx\\
&\approx-4\pi^2 (f_k-f_i) \left[\beta_2+\pi\beta_3(f_i+f_k\right)]f_1=\\
&=\phi_{i,k}f_1,
\label{appD:mismatch_term}
\end{split}
\end{equation}
with $\phi_{i,k} = -4\pi (f_k-f_i) \left[\beta_2+\pi\beta_3(f_i+f_k\right)]$. The most impacted channels by this approximation are the ones near the COI. The error relative to this approximation is given by \cite[Eq. 25]{danielclosed}.
 
Now, we consider Eq.~\eqref{eq:XPM_GN_integral} giving us the XPM contribution. For notation brevity, we will omit the factor $\frac{32}{27}\frac{\gamma^2}{B_k^2} \left( \frac{P_k}{P_i} \right)^2$. Also, the term $\Pi \left(\frac{f_1 + f_2}{B_k} \right)$ is neglected - this is equivalent to approximating the integration domain of the GN model to a rectangle~\cite{gnmodel}. Because of the approximation in Eq.~\eqref{appD:mismatch_term}, $\phi$ no longer depends on $f_2$, and the double integral in~\eqref{eq:XPM_GN_integral} turns to be a single integral. Thus, inserting Eq.~\eqref{eq:link_function_closed} in Eq.~\eqref{eq:XPM_GN_integral}, we can identify, three terms as follows

\begin{equation}
\begin{split}
&\eta_{\text{XPM}}^{(k)}(f_i) = \sum_{\substack{0 \leq l_1 + l_2 \leq 1 \\ 0 \leq l_1^\prime + l_2^\prime \leq 1}} \Upsilon_k \Upsilon_k^\prime  [(\kappa_{f,k} \kappa_{f,k}^\prime + \kappa_{b,k} \kappa_{b,k}^\prime) \times \\
& \times \eta_{\text{XPM,main}}^{(k)}(f_i) - (\kappa_{f,k} \kappa_{b,k}^\prime +  \kappa_{b,k} \kappa_{f,k}^\prime)\eta_{\text{XPM,cos}}^{(k)}(f_i) +\\
&+ (\kappa_{f,k} \kappa_{b,k}^\prime -  \kappa_{b,k} \kappa_{f,k}^\prime)\eta_{\text{XPM,sin}}^{(k)}(f_i)],
\label{appD:XPM_GN_integral1}
\end{split}
\end{equation}
with
\begin{equation}
\begin{split}
&\eta_{\text{XPM,main}}^{(k)}(f_i) = 2B_k\int_{0}^{\frac{B_i}{2}} df_1   \frac{\alpha_{l,k} \alpha_{l,k}^\prime + \phi_{i,k}^2f_1^2}{(\alpha_{l,k}^2 + \phi_{i,k}^2f_1^2)(\alpha_{l,k}^{\prime 2} + \phi_{i,k}^2f_1^2)},
\label{appD:XPM_GN_integral2}
\end{split}
\end{equation}
\begin{equation}
\begin{split}
&\eta_{\text{XPM,cos}}^{(k)}(f_i) = \\&= 2B_k\int_{0}^{\frac{B_i}{2}} df_1  \frac{\alpha_{l,k} \alpha_{l,k} ^\prime + \phi_{i,k}^2f_1^2}{(\alpha_{l,k}^2 + \phi_{i,k}^2f_1^2)(\alpha_{l,k}^{\prime 2} + \phi_{i,k}^2f_1^2)}\cos(\phi_{i,k} L f_1)
\label{appD:XPM_GN_integral3}
\end{split}
\end{equation}
and
\begin{equation}
\begin{split}
&\eta_{\text{XPM,sin}}^{(k)}(f_i)= \\&= 2B_k\int_{0}^{\frac{B_i}{2}} df_1  \frac{(\alpha_{l,k} - \alpha_{l,k}^\prime) \phi_{i,k}f_1 }{(\alpha_{l,k}^2 + \phi_{i,k}^2f_1^2)(\alpha_{l,k}^{\prime 2} + \phi_{i,k}^2f_1^2)}\sin(\phi_{i,k} L f_1).
\label{appD:XPM_GN_integral4}
\end{split}
\end{equation}
In the following, the above three integrals are solved. Eq.~\eqref{appD:XPM_GN_integral2} is solving using identity \eqref{appE:integral1} as
\begin{multline}
\eta_{\text{XPM,main}}^{(k)}(f_i) =\frac{2B_k}{\phi_{i,k}(\alpha_{l,k} + \alpha_{l,k}^\prime)}  \times\\
\times \left[\arctan\left(\frac{\phi_{i,k}B_i}{2\alpha_{l,k}}\right)
+ \arctan\left(\frac{\phi_{i,k}B_i}{2\alpha_{l,k}^\prime}\right)\right],
\label{appD:XPM_GN_integral5}
\end{multline}
Eqs.~\eqref{appD:XPM_GN_integral3} and \eqref{appD:XPM_GN_integral4} do not have analytical solutions in their current form. In order to derive an analytical solution, we extend the channel bandwidth $B_i \rightarrow \infty$ and solve it using identities~\eqref{appE:integral4} and~\eqref{appE:integral5}, yielding to
\begin{multline}
\eta_{\text{XPM,cos}}^{(k)}(f_i) =\frac{\pi B_k}{\phi_{i,k}(\alpha_{l,k} + \alpha_{l,k}^\prime)}  \times\\ 
\times \left[ e^{-|\alpha_{l,k} L|}\sign\left(\frac{\phi_{i,k}}{\alpha_{l,k}}\right) + e^{-|\alpha_{l,k}^\prime L|}\sign\left(\frac{\phi_{i,k}}{\alpha_{l,k}^\prime}\right) \right]
\label{appD:XPM_GN_integral6}
\end{multline}
and
\begin{multline}
\eta_{\text{XPM,sin}}^{(k)}(f_i) =\frac{\pi B_k}{\phi_{i,k}(\alpha_{l,k} + \alpha_{l,k}^\prime)}  \times \\
\times \left[ e^{-|\alpha_{l,k} L|}\sign\left(-\phi_{i,k}\right) + e^{-|\alpha_{l,k}^\prime L|}\sign\left(\phi_{i,k}\right) \right]
\label{appD:XPM_GN_integral7}
\end{multline}
Finally, by inserting Eqs.~\eqref{appD:XPM_GN_integral5}, \eqref{appD:XPM_GN_integral6} and \eqref{appD:XPM_GN_integral7} in Eq.\eqref{appD:XPM_GN_integral1} together with the pre-factor $\frac{32}{27}\frac{\gamma^2}{B_k^2} \left( \frac{P_k}{P_i} \right)^2$, Eq.~\eqref{eq:XPM_closed} is obtained concluding the proof.

\section{Derivation of the SPM contribution.}
\label{appD:SPM}

This section shows the derivation of Eq.~\eqref{eq:SPM_closed}. We start by approximating the phase mismatch term. We assume that the dispersion slope $\beta_3$ is constant over the channel bandwidth. Thus, the phase mismatch term can be approximated as  
\begin{equation}
\begin{split}
&\phi(f_1+f_i,f_2+f_i,f_i)=\\
&= -4\pi^2f_1f_2\left[\beta_2+\pi\beta_3(f_1+f_2-2f_i\right)] \approx\\
&\approx-4\pi^2f_1f_2(\beta_2+2\pi\beta_3f_i)=\\
&=\phi_{i}f_1f_2,
\label{appD:mismatch_term_SPM}
\end{split}
\end{equation}
with $\phi_{i} = -4\pi^2(\beta_2+2\pi\beta_3f_i)$. 

Now, using Eq.~\eqref{eq:SPM_GN_integral} together with Eqs.~\eqref{eq:XPM_GN_integral} and~\eqref{eq:link_function_closed} with $k=i$, and omitting the pre-factor of $\frac{16}{27}\frac{\gamma^2}{B_i^2}$, we can write
\begin{equation}
\begin{splitfit}
\eta_{\text{SPM}}(f_i) = \sum_{\substack{0 \leq l_1 + l_2 \leq 1 \\ 0 \leq l_1^\prime + l_2^\prime \leq 1}} &\Upsilon_i \Upsilon_i^\prime  \bigggl[(\kappa_{f,i} \kappa_{f,i}^\prime + \kappa_{b,i} \kappa_{b,i}^\prime)\eta_{\text{SPM,main}}(f_i)-\\
- &(\kappa_{f,i} \kappa_{b,i}^\prime +  \kappa_{b,i} \kappa_{f,i}^\prime)\eta_{\text{SPM,cos}}(f_i) +\\
+ &(\kappa_{f,i} \kappa_{b,i}^\prime -  \kappa_{b,i} \kappa_{f,i}^\prime)\eta_{\text{SPM,sin}}(f_i)\bigggr],
\label{appD:SPM_GN_integral1}
\end{splitfit}
\end{equation}
where $\eta_{\text{SPM}}(f_i)$, $\eta_{\text{SPM,cos}}(f_i)$ and $\eta_{\text{SPM,sin}}(f_i)$ are given respectively by

\begin{multline}
\eta_{\text{SPM,main}}(f_i) = \\ = \int_{-\frac{B_i}{2}}^{\frac{B_i}{2}} df_1
\int_{-\frac{B_i}{2}}^{\frac{B_i}{2}} df_2
\frac{\alpha_{l,i} \alpha_{l,i} ^\prime + \phi_i^2f_1^2f_2^2}{(\alpha_{l,i}^2 + \phi_i^2f_1^2f_2^2)(\alpha_{l,i}^{\prime 2} + \phi_i^2f_1^2f_2^2)},
\label{appD:SPM_GN_integral2}
\end{multline}
\begin{equation}
\begin{splitfit}
&\eta_{\text{SPM,cos}}(f_i) =\\&= \int_{-\frac{B_i}{2}}^{\frac{B_i}{2}} df_1  \int_{-\frac{B_i}{2}}^{\frac{B_i}{2}} df_2\frac{\alpha_{l,i} \alpha_{l,i} ^\prime + \phi_i^2f_1^2f_2^2}{(\alpha_{l,i}^2 + \phi_i^2f_1^2f_2^2)(\alpha_{l,i}^{\prime 2} + \phi_i^2f_1^2f_2^2)}\cos(\phi_i L f_1 f_2)
\label{appD:SPM_GN_integral3}
\end{splitfit}
\end{equation}
and
\begin{equation}
\begin{splitfit}
&\eta_{\text{SPM,sin}}(f_i) = \\&= \int_{-\frac{B_i}{2}}^{\frac{B_i}{2}} df_1 \int_{-\frac{B_i}{2}}^{\frac{B_i}{2}} df_2 \frac{(\alpha_{l,i} - \alpha_{l,i}^\prime) \phi_if_1f_2 }{(\alpha_{l,i}^2 + \phi_i^2f_1^2f_2^2)(\alpha_{l,i}^{\prime 2} + \phi_i^2f_1^2f_2^2)}\sin(\phi_i L f_1 f_2).
\label{appD:SPM_GN_integral4}
\end{splitfit}
\end{equation}
Note that, similar to Appendix~\ref{appC:XPM}, the term $\Pi \left(\frac{f_1 + f_2}{B_i} \right)$ is neglected.

In the following, the three integrals above are solved. The integral in Eq.~\eqref{appD:SPM_GN_integral2} is rewritten in polar coordinates $(r,\varphi)$ as
\begin{equation}
\begin{split}
&\eta_{\text{SPM,main}}(f_i) \approx 4\int_{0}^{\sqrt{\frac{3}{\pi}}\frac{B_i}{2}} dr \int_{0}^{\frac{\pi}{2}} d\varphi  \ \times \\
&\times \frac{r\left[\alpha_{l,i} \alpha_{l,i}^\prime + \frac{\phi_i^2}{4} (r^4\sin^2{(\varphi)}) \right]}{\left[\alpha_{l,i}^2 + \frac{\phi_i^2}{4} (r^4\sin^2{(\varphi)}) \right]\left[\alpha_{l,i}^{\prime 2} + \frac{\phi_i^2}{4} (r^4\sin^2{(\varphi)}) \right]},
\label{appD:SPM_GN_integral5}
\end{split}
\end{equation}
where it was used the relations $f_1 = r\cos{(\varphi /2)}$, $f_2 = r\sin{(\varphi / 2)}$ and $ \sin{(\varphi / 2)}\cos{(\varphi / 2)} = \frac{\sin{(\varphi)}}{2}$. Also, the integration domain of Eq.~\eqref{eq:SPM_GN_integral} was approximated by a circular domain such that the area of both domains is equal~\cite[Fig. 3]{danielclosed}. This yields the variation of the radius in the outer integral as shown in Eq.~\eqref{appD:SPM_GN_integral5}. 
The inner integral in Eq.~\eqref{appD:SPM_GN_integral5} can be solved using identity \eqref{appE:integral2}, yielding to
\begin{equation}
\begin{split}
&\eta_{\text{SPM,main}}(f_i) \approx 4\int_{0}^{\sqrt{\frac{3}{\pi}}\frac{B_i}{2}} dr\times\\
&\times  \frac{ r \pi}{\alpha_{l,i} +\alpha_{l,i}^\prime }\left[\frac{1}{\sqrt{4 \alpha_{l,i}^2 + \phi_i^2 r^4}} + \frac{1}{\sqrt{4 \alpha_{l,i}^{\prime 2} + \phi_i^2 r^4}}\right].
\label{appD:SPM_GN_integral6}
\end{split}
\end{equation}
This integral can be rewritten as:
\begin{multline}
\eta_{\text{SPM,main}}(f_i) =  \frac{ 2\pi}{\alpha_{l,i} +\alpha_{l,i}^\prime }\int_{0}^{\sqrt{\frac{3}{\pi}}\frac{B_i}{2}} dr\times\\
\times \left[\frac{r}{ \alpha_{l,i}\sqrt{1 + \frac{\phi_i^2 r^4}{4 \alpha_{l,i}^2}}} + \frac{r}{ \alpha_{l,i}^\prime\sqrt{1 + \frac{\phi_i^2 r^4}{4 \alpha_{l,i}^{\prime 2}}}}\right].
\label{appD:SPM_GN_integral7}
\end{multline}
The integral in Eq.~\eqref{appD:SPM_GN_integral7} is solved using identity \eqref{appE:integral3} as
\begin{multline}
\eta_{\text{SPM,main}}(f_i) =  \frac{ 2\pi}{\phi_i(\alpha_{l,i} +\alpha_{l,i}^\prime)} \times\\
\times
\left[\asinh{\left(\frac{3 \phi_i B_i^2}{8 \pi \alpha_{l,i} } \right)} + \asinh{\left(\frac{3 \phi_i B_i^2}{8 \pi \alpha_{l,i}^\prime } \right)} \right].
\label{appD:SPM_GN_integral8}
\end{multline}

To solve the integrals in Eqs.~\eqref{appD:SPM_GN_integral3} and~\eqref{appD:SPM_GN_integral4}, a similar approach use in~\cite{RamanDan} is used. The integrals are converted to hyperbolic coordinates using the relations $\nu_1 = \sqrt{f_1 f_2}$, $\nu_2 = -\frac{1}{2} \ln \left(\frac{f_1}{f_2}\right)$, $f_1 = \nu_1 e^{\nu_2}$ and $f_2 = \nu_1 e^{-\nu_2}$~\cite[Sec.~VIII-A]{gnmodel}; this change of coordinates yields a one-dimensional integral in $\nu_1$. We also use the change of variable $\nu = \nu_1^2$~\cite{RamanDan} to rewrite Eqs.~\eqref{appD:SPM_GN_integral3} and \eqref{appD:SPM_GN_integral4} as
\begin{equation}
\begin{splitfit}
&\eta_{\text{SPM,cos}}(f_i) = \\&=8\int_{0}^{\frac{B_i}{2}} d\nu \ln\left(\frac{B_i}{2\sqrt{\nu}}\right) \frac{\alpha_{l,i} \alpha_{l,i}^\prime + \phi_i^2\nu^2}{(\alpha_{l,i}^2 + \phi_i^2\nu^2)(\alpha_{l,i}^{\prime 2} + \phi_i^2\nu^2)}\cos(\phi_i L \nu)
\label{appD:SPM_GN_integral9}
\end{splitfit}
\end{equation}
and
\begin{equation}
\begin{splitfit}
&\eta_{\text{SPM,sin}}(f_i) =\\ & = 8\int_{0}^{\frac{B_i}{2}} d\nu \ln\left(\frac{B_i}{2\sqrt{\nu}}\right) \frac{(\alpha_{l,i} - \alpha_{l,i} ^\prime) \phi_i\nu}{(\alpha_{l,i}^2 + \phi_i^2\nu^2)(\alpha_{l,i}^{\prime 2} + \phi_i^2\nu^2)}\sin(\phi_i L \nu).
\label{appD:SPM_GN_integral10}
\end{splitfit}
\end{equation}
The integrals in Eqs.~\eqref{appD:SPM_GN_integral9} and \eqref{appD:SPM_GN_integral10} do not have analytical solutions in their current form. In order to obtain an integral that yields an analytical solution we evaluate the logarithm functions in the point $\nu = \frac{\pi}{2\phi_i L}$, where this point was chosen such that the cosine function achieves its minima and the sine function achieves its maxima. This yields to  
\begin{equation}
\begin{splitfit}
&\eta_{\text{SPM,cos}}(f_i) = \\& = 8\ln\left(\sqrt{\frac{\phi_i L}{2 \pi}}B_i\right)\int_{0}^{\frac{B_i}{2}} d\nu \frac{\alpha_{l,i} \alpha_{l,i} ^\prime + \phi_i^2\nu^2}{(\alpha_{l,i}^2 + \phi_i^2\nu^2)(\alpha_{l,i}^{\prime 2} + \phi_i^2\nu^2)}\cos(\phi_i L \nu)
\label{appD:SPM_GN_integral11}
\end{splitfit}
\end{equation}
and
\begin{equation}
\begin{splitfit}
&\eta_{\text{SPM,sin}}(f_i) =\\ & = 8\ln\left(\sqrt{\frac{\phi_i L}{2 \pi}}B_i\right)\int_{0}^{\frac{B_i}{2}} d\nu  \frac{(\alpha_{l,i} - \alpha_{l,i} ^\prime) \phi_i\nu}{(\alpha_{l,i}^2 + \phi_i^2\nu^2)(\alpha_{l,i}^{\prime 2} + \phi_i^2\nu^2)}\sin(\phi_i L \nu).
\label{appD:SPM_GN_integral12}
\end{splitfit}
\end{equation}
The integrals in Eqs.~\eqref{appD:SPM_GN_integral11} and \eqref{appD:SPM_GN_integral12} can now be solved similar to Appendix~\ref{appC:XPM}, i.e., by letting $B_i \rightarrow \infty$. This yields to 
\begin{multline}
\eta_{\text{SPM,cos}}(f_i) = 4\pi\ln\left(\sqrt{\frac{\phi_i L}{2 \pi}}B_i\right) \times \\
\times \left[ e^{-|\alpha_{l,i} L|}\sign\left(\frac{\phi_i}{\alpha_{l,i}}\right) + e^{-|\alpha_{l,i}^\prime L|}\sign\left(\frac{\phi_i}{\alpha_{l,i}^\prime}\right) \right]
\label{appD:SPM_GN_integral13}
\end{multline}
and
\begin{multline}
\eta_{\text{SPM,sin}}(f_i) = 4\pi\ln\left(\sqrt{\frac{\phi_i L}{2 \pi}}B_i\right) \times \\ 
\times \left[ e^{-|\alpha_{l,i} L|}\sign\left(-\phi_i\right) + e^{-|\alpha_{l,i}^\prime L|}\sign\left(\phi_i\right) \right].
\label{appD:SPM_GN_integral14}
\end{multline}
Finally, by inserting Eqs.~\eqref{appD:SPM_GN_integral8}, \eqref{appD:SPM_GN_integral13} and \eqref{appD:SPM_GN_integral14} in Eq.~\eqref{appD:SPM_GN_integral1} together with the pre-factor of  $\frac{16}{27}\frac{\gamma^2}{B_i^2}$, Eq.~\eqref{eq:SPM_closed} is obtained concluding the proof.

\section{Mathematical Identities}
\label{appE:Mathematical_identities}


\begin{multline}
(x + y + z)^i = \\= \sum_{0 \leq 1_1 + 1_2 \leq i} \frac{i!}{l_1! l_2!(i-l_1-1_2)!} x^{l_1}y^{l_2}z^{i-l_1-l_2}.
\label{appE:multinomial_theorem}
\end{multline}

\begin{equation}
\begin{split}
|z_k|^2 = \Re{( z_k \cdot \overline{z}_k)} = z_k \cdot \overline{z}_k. 
\label{appE:modulo_complex_number}
\end{split}
\end{equation}

\begin{equation}
\begin{split}
z_i \cdot \overline{z}_j +  z_j \cdot \overline{z}_i  = 2\Re{( z_i \cdot \overline{z}_j)}, \text{    } j<i.
\label{appE:sum_modulo_complex_number}
\end{split}
\end{equation}

\begin{equation}
\begin{split}
&\int_{0}^{X}dx \ \frac{ab + c^2x^2}{(a^2 + c^2x^2)(b^2 + c^2x^2)}= \\
&=\frac{1}{c(a+b)}\left[\arctan\left(\frac{cx}{a}\right) + \arctan\left(\frac{cx}{b}\right)\right].
\label{appE:integral1}
\end{split}
\end{equation}

\begin{equation}
\begin{split}
&\int_{0}^{\frac{\pi}{2}}dx \ \frac{ab + c^2\sin^2{(x)}}{[a^2 + c^2\sin^2{(x)}][b^2 + c^2\sin^2{(x)}]}= \\
&= \frac{\pi}{2(a+b)}\left(\frac{1}{\sqrt{a^2 + c^2}} + \frac{1}{\sqrt{b^2 +c^2 }}\right).
\label{appE:integral2}
\end{split}
\end{equation}

\begin{equation}
\begin{split}
&\int_{0}^{X}dx \ \frac{x}{\sqrt{1 + d^2x^4}}
= \frac{1}{2d}\asinh{(dX^2)}.
\label{appE:integral3}
\end{split}
\end{equation}

\begin{equation}
\begin{split}
&\int_{0}^{\infty}dx \ \frac{ab + c^2x^2}{(a^2 + c^2x^2)  (b^2 + c^2x^2)} \cos(cxL) =\\
&= \frac{\pi}{2} \frac{e^{-|aL|}\sign(c/a) + e^{-|bL|}\sign(c/b)}{c(a + b)}.
\label{appE:integral4}
\end{split}
\end{equation}

\begin{equation}
\begin{split}
&\int_{0}^{\infty}dx \ \frac{(a-b)cx}{(a^2 + c^2x^2)  (b^2 + c^2x^2)}\sin(cxL)  =\\
&= \frac{\pi}{2} \frac{e^{-|aL|}\sign(-c) + e^{-|bL|}\sign(c)}{c(a + b)}.
\label{appE:integral5}
\end{split}
\end{equation}

\bibliographystyle{IEEEbib}
\bibliography{IEEEabrv,main}

\end{document}

%% file: Figures/loss_and_raman.tex
\begin{tikzpicture}[baseline]
\begin{axis}[yshift=-0.2cm,
legend cell align=left,
title={\footnotesize (a)},
legend style={font=\footnotesize, at={(rel axis cs:0,1)}, anchor=north west},
width=\linewidth, height = 4.8cm,
xlabel={Wavelength [nm]},
ylabel={Attenuation [dB/km]},
grid=both,
ymax=0.28,ymin=0.18,
xmin=1400,xmax=1620,
xticklabel style={/pgf/number format/1000 sep=},
ytick distance=0.02,
xtick distance=40,
]

\addplot[black,very thick] table[x=wavelength,y=loss] {data/attenuation_profile.txt};

\node[anchor=west] (source) at (axis cs:1428-6,0.27){E};
\node[anchor=west] (source) at (axis cs:1493-6,0.27){S};
\node[anchor=west] (source) at (axis cs:1543-6,0.27){C};
\node[anchor=west] (source) at (axis cs:1588-6,0.27){L};

\begin{scope}[on background layer]
    \fill[red,opacity=0.1] ({rel axis cs:0.0,0.0}) rectangle ({rel axis cs:0.2727,1});
    \fill[green,opacity=0.1] ({rel axis cs:0.2727,0}) 
    rectangle ({rel axis cs:0.5909,1});
    \fill[blue,opacity=0.1] ({rel axis cs:0.5909,0})
    rectangle ({rel axis cs:0.75,1});
    \fill[gray,opacity=0.1] ({rel axis cs:0.75,0})
    rectangle ({rel axis cs:1,1});
\end{scope}

\end{axis}




\end{tikzpicture}
\begin{tikzpicture}[baseline]
\begin{axis}[yshift=-0.2cm,
title={\footnotesize (b)},
legend cell align=left,
legend style={font=\footnotesize, at={(rel axis cs:0,1)}, anchor=north west},
width=\linewidth, height = 4.8cm,
xlabel={Frequency separation [THz]},
ylabel={Raman gain [1/km/W]},
grid=both,
ymax=0.5,ymin=0,
xmin=0,xmax=26,
    ytick distance=0.1,
    xtick distance=4,
]

\addplot[Set1-C,thick] table[x=Delta_Freq,y=Raman_Gain] {data/JLT_Raman.txt};

\end{axis}
\end{tikzpicture}

%% file: Figures/pumps.tex
\begin{tabular}{cccccccccccc}
\multicolumn{1}{l}{}                                             & \multicolumn{9}{c}{E-band}
& \multicolumn{2}{c}{S-band}                                     \\ \hline

\multicolumn{1}{|c|}{\cellcolor[HTML]{E0E0E0}Wavelength {[}nm{]}}      
& \multicolumn{1}{c|}{\cellcolor[HTML]{FFD0D0}1402.1} 
& \multicolumn{1}{c|}{\cellcolor[HTML]{FFD0D0}1408.7} 
& \multicolumn{1}{c|}{\cellcolor[HTML]{FFD0D0}1415.4} 
& \multicolumn{1}{c|}{\cellcolor[HTML]{FFD0D0}1422.1} 
& \multicolumn{1}{c|}{\cellcolor[HTML]{FFD0D0}1428.8} 
& \multicolumn{1}{c|}{\cellcolor[HTML]{FFD0D0}1435.7} 
& \multicolumn{1}{c|}{\cellcolor[HTML]{FFD0D0}1442.6} 
& \multicolumn{1}{c|}{\cellcolor[HTML]{FFD0D0}1449.6} 
& \multicolumn{1}{c|}{\cellcolor[HTML]{FFD0D0}1456.6} 
& \multicolumn{1}{c|}{\cellcolor[HTML]{D5F6D5}1463.7} 
& \multicolumn{1}{c|}{\cellcolor[HTML]{D5F6D5}1485.4} 
\\ \hline

\multicolumn{12}{|c|}{Forward Raman Pump Scenario} 
\\ \hline

\multicolumn{1}{|c|}{\cellcolor[HTML]{E0E0E0}FW pumps' power at $z = 0$~{[}mW{]}} 
& \multicolumn{1}{c|}{\cellcolor[HTML]{FFD0D0}150.9}                          
& \multicolumn{1}{c|}{\cellcolor[HTML]{FFD0D0}331.3}                          
& \multicolumn{1}{c|}{\cellcolor[HTML]{FFD0D0}161.2}                          
& \multicolumn{1}{c|}{\cellcolor[HTML]{FFD0D0}119.5}                          
& \multicolumn{1}{c|}{\cellcolor[HTML]{FFD0D0}34.3}                           
& \multicolumn{1}{c|}{\cellcolor[HTML]{FFD0D0}35.8}                           
& \multicolumn{1}{c|}{\cellcolor[HTML]{FFD0D0}30.4}                           
& \multicolumn{1}{c|}{\cellcolor[HTML]{FFD0D0}25.7}
& \multicolumn{1}{c|}{-}
& \multicolumn{1}{c|}{\cellcolor[HTML]{D5F6D5}63.0}                           
& \multicolumn{1}{c|}{\cellcolor[HTML]{D5F6D5}17.2}                           
\\ \hline

\multicolumn{12}{|c|}{Backward Raman Pump Scenario} 
\\ \hline

\multicolumn{1}{|c|}{\cellcolor[HTML]{E0E0E0}BW pumps' power at $z = L$ ~{[}mW{]}} 
& \multicolumn{1}{c|}{-}                              
& \multicolumn{1}{c|}{\cellcolor[HTML]{FFD0D0}668.7}                          
& \multicolumn{1}{c|}{\cellcolor[HTML]{FFD0D0}64.6}                           
& \multicolumn{1}{c|}{\cellcolor[HTML]{FFD0D0}167.7}                          
& \multicolumn{1}{c|}{\cellcolor[HTML]{FFD0D0}14.3}                           
& \multicolumn{1}{c|}{\cellcolor[HTML]{FFD0D0}58.2}                           
& \multicolumn{1}{c|}{\cellcolor[HTML]{FFD0D0}45.3}                           
& \multicolumn{1}{c|}{\cellcolor[HTML]{FFD0D0}50.8}                           
& \multicolumn{1}{c|}{-}
& \multicolumn{1}{c|}{\cellcolor[HTML]{D5F6D5}13.4}                           
& \multicolumn{1}{c|}{\cellcolor[HTML]{D5F6D5}58.5}                           
\\ \hline

\multicolumn{1}{|c|}{\cellcolor[HTML]{E0E0E0}BW pumps' power at $z=0$~ {[}µW{]}}
    &  \multicolumn{1}{c|}{-}
    &  \multicolumn{1}{c|}{\cellcolor[HTML]{FFD0D0}203.9}
    &  \multicolumn{1}{c|}{\cellcolor[HTML]{FFD0D0}40.0}
    &  \multicolumn{1}{c|}{\cellcolor[HTML]{FFD0D0}200.9}
    &  \multicolumn{1}{c|}{\cellcolor[HTML]{FFD0D0}30.7}
    &  \multicolumn{1}{c|}{\cellcolor[HTML]{FFD0D0}198.7}
    &  \multicolumn{1}{c|}{\cellcolor[HTML]{FFD0D0}225.6}
    &  \multicolumn{1}{c|}{\cellcolor[HTML]{FFD0D0}350.0}
    &  \multicolumn{1}{c|}{-}
    &  \multicolumn{1}{c|}{\cellcolor[HTML]{D5F6D5}186.4}
    &  \multicolumn{1}{c|}{\cellcolor[HTML]{D5F6D5}5235}
\\ \hline

\multicolumn{12}{|c|}{Forward + Backward Raman Pump Scenario} 
\\ \hline
\multicolumn{1}{|c|}{\cellcolor[HTML]{E0E0E0}FW pump power at $z = 0$~{[}mW{]}}
    & \multicolumn{1}{c|}{-}
    & \multicolumn{1}{c|}{-}
    &  \multicolumn{1}{c|}{-}
    &  \multicolumn{1}{c|}{-}
    &  \multicolumn{1}{c|}{-}
    &  \multicolumn{1}{c|}{-}
    &  \multicolumn{1}{c|}{-}
    &  \multicolumn{1}{c|}{-}
    &  \multicolumn{1}{c|}{-}
    &  \multicolumn{1}{c|}{-}
    &  \multicolumn{1}{c|}{\cellcolor[HTML]{D5F6D5}393.32}
\\ \hline

\multicolumn{1}{|c|}{\cellcolor[HTML]{E0E0E0}BW pumps' power at $z = L$~ {[}mW{]}}
    &  \multicolumn{1}{c|}{\cellcolor[HTML]{FFD0D0}297.79}
    &  \multicolumn{1}{c|}{\cellcolor[HTML]{FFD0D0}123.07}
    &  \multicolumn{1}{c|}{\cellcolor[HTML]{FFD0D0}130.92}
    &  \multicolumn{1}{c|}{\cellcolor[HTML]{FFD0D0}184.78}
    &  \multicolumn{1}{c|}{-}
    &  \multicolumn{1}{c|}{\cellcolor[HTML]{FFD0D0}80.68}
    &  \multicolumn{1}{c|}{\cellcolor[HTML]{FFD0D0}17.88}
    &  \multicolumn{1}{c|}{-}
    &  \multicolumn{1}{c|}{\cellcolor[HTML]{FFD0D0}24.23}
    &  \multicolumn{1}{c|}{\cellcolor[HTML]{D5F6D5}27.41}
    &  \multicolumn{1}{c|}{-}
\\ \hline

\multicolumn{1}{|c|}{\cellcolor[HTML]{E0E0E0}BW pumps' power at $z=0$~ {[}µW{]}}
    &  \multicolumn{1}{c|}{\cellcolor[HTML]{FFD0D0}53.9}
    &  \multicolumn{1}{c|}{\cellcolor[HTML]{FFD0D0}43.7}
    &  \multicolumn{1}{c|}{\cellcolor[HTML]{FFD0D0}96.6}
    &  \multicolumn{1}{c|}{\cellcolor[HTML]{FFD0D0}263.5}
    &  \multicolumn{1}{c|}{-}
    &  \multicolumn{1}{c|}{\cellcolor[HTML]{FFD0D0}327.9}
    &  \multicolumn{1}{c|}{\cellcolor[HTML]{FFD0D0}103.9}
    &  \multicolumn{1}{c|}{-}
    &  \multicolumn{1}{c|}{\cellcolor[HTML]{FFD0D0}250.8}
    &  \multicolumn{1}{c|}{\cellcolor[HTML]{D5F6D5}396.7}
    &  \multicolumn{1}{c|}{-}
\\ \hline

\end{tabular}

%% file: Figures/power_profile.tex
  \begin{tikzpicture}[baseline]
    \begin{axis}[
    unbounded coords=jump,view={50}{20}, grid=both,
    legend cell align=left,title={(a)},
legend style={font=\footnotesize, at={(rel axis cs:0.5,0.6,1.1)}, anchor=north},
width=7.5cm,
    x label style={rotate=-20},
     y label style={rotate=15},
    xlabel={Distance [km]},
ylabel={Wavelength [nm]},
zlabel={Power profile [dBm]},
x tick label style = {text width = 1.0cm, align = center, rotate = 70},
xtick={0,20,40,60,80},
ytick={1500,1550,1600},
zmin=-14,zmax=6,
ymin=1490,ymax=1610,
xmin=0,xmax=80,
xticklabel style={/pgf/number format/1000 sep=},
yticklabel style={/pgf/number format/1000 sep=},
  ztick distance=2,
    ]
    
      \addplot3[surf,color=green,opacity=0.7] file {data/JLT_power_forward.txt};
      \addlegendentry{Forward Raman Amplification}
    \end{axis}
\end{tikzpicture}
  \begin{tikzpicture}[baseline]
    \begin{axis}[
    unbounded coords=jump,view={50}{20}, grid=both,
    legend cell align=left,title={(b)},
legend style={font=\footnotesize, at={(rel axis cs:0.5,0.6,1.1)}, anchor=north},
width=7.5cm,
    x label style={rotate=-20},
     y label style={rotate=15},
    xlabel={Distance [km]},
ylabel={Wavelength [nm]},
zlabel={Power profile [dBm]},
x tick label style = {text width = 1.0cm, align = center, rotate = 70},
xtick={0,20,40,60,80},
ytick={1500,1550,1600},
zmin=-14,zmax=6,
ymin=1490,ymax=1610,
xmin=0,xmax=80,
xticklabel style={/pgf/number format/1000 sep=},
yticklabel style={/pgf/number format/1000 sep=},
  ytick distance=134,
  ztick distance=2,
    ]
    
      \addplot3[surf,color=blue,opacity=0.4] file {data/JLT_power_backward.txt};
      \addlegendentry{Backward Raman Amplification}
    \end{axis}
\end{tikzpicture}
  \begin{tikzpicture}[baseline]
    \begin{axis}[
    unbounded coords=jump,view={50}{20}, grid=both,
    legend cell align=left,title={(c)},
legend style={font=\footnotesize, at={(rel axis cs:0.5,0.6,1.12)}, anchor=north},
width=7.5cm,
    x label style={rotate=-20},
     y label style={rotate=15},
    xlabel={Distance [km]},
ylabel={Wavelength [nm]},
zlabel={Power profile [dBm]},
x tick label style = {text width = 1.0cm, align = center, rotate = 70},
xtick={0,20,40,60,80},
ytick={1500,1550,1600},
zmin=-14,zmax=8,
ymin=1490,ymax=1610,
xmin=0,xmax=80,
xticklabel style={/pgf/number format/1000 sep=},
yticklabel style={/pgf/number format/1000 sep=},
  ytick distance=134,
  ztick distance=2,
    ]
    
      \addplot3[surf,color=red,opacity=0.4] file {data/JLT_power_forward_and_backward.txt};
      \addlegendentry{Forward + Backward Raman Amplification}
    \end{axis}
\end{tikzpicture}

%% file: Figures/fitting.tex
\pgfplotsset{simulation/.style={very thick}}
\pgfplotsset{integral/.style={thick}}
\pgfplotsset{closedform/.style={thick}}

\pgfplotsset{dual-line/.style 2 args={legend image code/.code={
      \draw[#1] plot coordinates { (0cm,.05cm) (.6cm,.05cm) };%
      \draw[#2] plot coordinates { (0cm,-.05cm) (.6cm,-.05cm) };%
}}}

\begin{multicols}{2}

\begin{tikzpicture}[baseline]
\begin{axis}[
unbounded coords=jump,
title = (a) Worst fitting channel for FW and FW+BW scenarios,
legend style={font=\footnotesize, at={(rel axis cs:0,0)}, 
anchor=south west},
width=\linewidth, height = 5.1cm,
xlabel={Distance [km]},
ylabel={Normalised power [dB]},
grid=both,
ymax=9,ymin=-5,
xmin=0,xmax=80,
xticklabel style={/pgf/number format/1000 sep=},
/pgf/number format/fixed,
  ylabel near ticks,
    ytick distance=2,
    xtick distance=10,
]

\draw [black] \pgfextra{
    \pgfpathellipse{\pgfplotspointaxisxy{30}{3}}
    {\pgfplotspointaxisdirectionxy{4}{0}}
    {\pgfplotspointaxisdirectionxy{0}{1.5}}
}; 

\node[anchor=west] (source) at (axis cs:27,2.5){};
\node (destination) at (axis cs:17,1){};
\draw[->](source)--(destination);
\node[] at (axis cs: 15,0) {FW+BW}; 

\draw [black] \pgfextra{
    \pgfpathellipse{\pgfplotspointaxisxy{45}{3.8}}
    {\pgfplotspointaxisdirectionxy{4}{0}}
    {\pgfplotspointaxisdirectionxy{0}{1.5}}
}; 

\node[anchor=east] (source) at (axis cs:48,4.8){};
\node (destination) at (axis cs:56,5.8){};
\draw[->](source)--(destination);
\node[] at (axis cs: 64,5.8) {FW Only};

\addlegendimage{dual-line={Set1-C,thick}{Set1-A,thick}}
\addlegendentry{Numerical solution Eq.~\eqref{eq:diff_Raman}}
\addlegendimage{dual-line={Set1-C,thick,dashed}{Set1-A,thick,dashed}}
\addlegendentry{Semi-analytical solution Eq.~\eqref{eq:Raman_taylor}}

\addplot[Set1-C,integral] table[x=distance,y=solution] {data/fitting_FW.txt};
\addplot[Set1-A,integral] table[x=distance,y=solution] {data/fitting_FWBW.txt};
\addplot[Set1-C,integral, dashed] table[x=distance,y=fitting] {data/fitting_FW.txt};
\addplot[Set1-A,integral, dashed] table[x=distance,y=fitting] {data/fitting_FWBW.txt};

\end{axis}
\end{tikzpicture}

\columnbreak

\begin{tikzpicture}[baseline]
\begin{axis}[
unbounded coords=jump,
title = (b) Worst fitting channel for BW scenario,
legend style={font=\footnotesize, at={(rel axis cs:0.55,1)}, anchor=north},
width=\linewidth, height = 5.1cm,
xlabel={Distance [km]},
ylabel={Normalised power [dB]},
grid=both,
ymax=2,ymin=-14,
xmin=0,xmax=80,
xticklabel style={/pgf/number format/1000 sep=},
/pgf/number format/fixed,
  ylabel near ticks,
    ytick distance=2,
    xtick distance=10,
]
\addlegendentry{Numerical solution Eq.~\eqref{eq:diff_Raman}}
\addplot[Set1-B,integral] table[x=distance,y=solution] {data/fitting_BW.txt};
\addlegendentry{Semi-analytical solution Eq.~\eqref{eq:Raman_taylor}}
\addplot[Set1-B,integral, dashed] table[x=distance,y=fitting] {data/fitting_BW.txt};

\end{axis}
\end{tikzpicture}

\end{multicols}

%% file: Figures/SNR.tex
\pgfplotsset{simulation/.style={very thick}}
\pgfplotsset{integral/.style={thick}}
\pgfplotsset{closedform/.style={thick}}

\begin{tikzpicture}[baseline]
\begin{axis}[
unbounded coords=jump,
title = (a) Forward Raman Amplification,
legend style={font=\footnotesize, at={(rel axis cs:1,0)}, anchor=south east},
width=\linewidth, height = 6cm,
xlabel={Wavelength [nm]},
ylabel={$\text{SNR}_{\text{NLI}}$ [dB]},
grid=both,
xtick={1490,1510,1530,1550,1570,1590,1610},
ytick={14,18,22,26,30,34,38},
ymax=38,ymin=14,
xmin=1490,xmax=1610,
xticklabel style={/pgf/number format/1000 sep=},
  ylabel near ticks,
    ytick distance=4,
    xtick distance=20,
]

\node[anchor=west] (source) at (axis cs:1510,36.8){S};
\node[anchor=west] (source) at (axis cs:1545,36.8){C};
\node[anchor=west] (source) at (axis cs:1580,36.8){L};

           \draw [black] \pgfextra{
\pgfpathellipse{\pgfplotspointaxisxy{1550}{32.7}}
{\pgfplotspointaxisdirectionxy{5}{0}}
{\pgfplotspointaxisdirectionxy{0}{2}}
}; 

\node[anchor=west] (source) at (axis cs:1550,31.9){};
\node (destination) at (axis cs:1563,31.9){};
\draw[->](source)--(destination);

\node[] at (axis cs: 1577,31.9) {Single span}; 

           \draw [black] \pgfextra{
\pgfpathellipse{\pgfplotspointaxisxy{1549}{27.7}}
{\pgfplotspointaxisdirectionxy{5}{0}}
{\pgfplotspointaxisdirectionxy{0}{2}}
}; 
\node[] at (axis cs: 1570,26.2) {3 spans}; 

\node[anchor=west] (source) at (axis cs:1549,26.7){};
\node (destination) at (axis cs:1561.5,26.2){};
\draw[->](source)--(destination);

           \draw [black] \pgfextra{
\pgfpathellipse{\pgfplotspointaxisxy{1549}{22}}
{\pgfplotspointaxisdirectionxy{5}{0}}
{\pgfplotspointaxisdirectionxy{0}{2}}
}; 
\node[] at (axis cs: 1540,17.2) {10 spans}; 

\node[anchor=west] (source) at (axis cs:1546,21.0){};
\node (destination) at (axis cs:1540,17.6){};
\draw[->](source)--(destination);

\begin{scope}[on background layer]
    \fill[green,opacity=0.1] ({rel axis cs:0.0,0.0}) rectangle ({rel axis cs:0.3,1});
    \fill[blue,opacity=0.1] ({rel axis cs:0.3,0}) 
    rectangle ({rel axis cs:0.65,1});
    \fill[gray,opacity=0.1] ({rel axis cs:0.65,0})
    rectangle ({rel axis cs:1,1});
\end{scope}

\addlegendentry{Closed-form}
\addplot[Set1-B,closedform] table[x=wav,y=fw1] {data/closed_form.txt};
\addplot[Set1-B,closedform,forget plot] table[x=wav,y=fw3] {data/closed_form.txt};
\addplot[Set1-B,closedform,forget plot] table[x=wav,y=fw10] {data/closed_form.txt};

\addlegendentry{Integral}
\addplot[Set1-C,integral,dashed] table[x=wav,y=fw1] {data/integral.txt};
\addplot[Set1-C,integral, dashed,forget plot] table[x=wav,y=fw3] {data/integral.txt};
\addplot[Set1-C,integral, dashed,forget plot] table[x=wav,y=fw10] {data/integral.txt};

\addlegendentry{Simulation}
\addplot[Set1-A,simulation,dotted] table[x=wav,y=fw1]
{data/SSFM.txt};
\addplot[Set1-A,simulation,dotted,forget plot] table[x=wav,y=fw3] {data/SSFM.txt};
\addplot[Set1-A,simulation,dotted,forget plot] table[x=wav,y=fw10] {data/SSFM.txt};

\end{axis}
\end{tikzpicture}
\begin{tikzpicture}[baseline]
\begin{axis}[
unbounded coords=jump,
title = (b) Backward Raman Amplification,
legend style={font=\footnotesize, at={(rel axis cs:0,0)}, anchor=south west},
 legend columns=3,
width=\linewidth, height = 6cm,
xlabel={Wavelength [nm]},
ylabel={$\text{SNR}_{\text{NLI}}$ [dB]},
grid=both,
xtick={1490,1510,1530,1550,1570,1590,1610},
ytick={20,24,28,32,36,40},
ymax=40,ymin=20,
xmin=1490,xmax=1610,
xticklabel style={/pgf/number format/1000 sep=},
/pgf/number format/fixed,
  ylabel near ticks,
    ytick distance=4,
    xtick distance=20,
]
\begin{scope}[on background layer]
    \fill[green,opacity=0.1] ({rel axis cs:0.0,0.0}) rectangle ({rel axis cs:0.3,1});
    \fill[blue,opacity=0.1] ({rel axis cs:0.3,0}) 
    rectangle ({rel axis cs:0.65,1});
    \fill[gray,opacity=0.1] ({rel axis cs:0.65,0})
    rectangle ({rel axis cs:1,1});
\end{scope}
    
\node[anchor=west] (source) at (axis cs:1510,39){S};
\node[anchor=west] (source) at (axis cs:1545,39){C};
\node[anchor=west] (source) at (axis cs:1580,39){L};

\draw [black] \pgfextra{
    \pgfpathellipse{\pgfplotspointaxisxy{1550}{35.7}}
    {\pgfplotspointaxisdirectionxy{5}{0}}
    {\pgfplotspointaxisdirectionxy{0}{2}}
}; 

\node[anchor=west] (source) at (axis cs:1549,34.4){};
\node (destination) at (axis cs:1563,33.3){};
\draw[->](source)--(destination);

\node[] at (axis cs: 1577,33.3) {Single span}; 

\draw [black] \pgfextra{
    \pgfpathellipse{\pgfplotspointaxisxy{1550}{30.5}}
    {\pgfplotspointaxisdirectionxy{5}{0}}
    {\pgfplotspointaxisdirectionxy{0}{2}}
}; 
\node[] at (axis cs: 1573,27.7) {3 spans}; 

\node[anchor=east] (source) at (axis cs:1553,29.2){};
\node (destination) at (axis cs:1564,27.7){};
\draw[->](source)--(destination);

\draw [black] \pgfextra{
    \pgfpathellipse{\pgfplotspointaxisxy{1550}{25.5}}
    {\pgfplotspointaxisdirectionxy{5}{0}}
    {\pgfplotspointaxisdirectionxy{0}{2}}
}; 
\node[] at (axis cs: 1527,23.7) {10 spans}; 

\node[anchor=east] (source) at (axis cs:1551,24){};
\node (destination) at (axis cs:1537.5,23.7){};
\draw[->](source)--(destination);

\addlegendentry{Closed-form}
\addplot[Set1-B,closedform] table[x=wav,y=bw1] {data/closed_form.txt};
\addplot[Set1-B,closedform,forget plot] table[x=wav,y=bw3] {data/closed_form.txt};
\addplot[Set1-B,closedform,forget plot] table[x=wav,y=bw10] {data/closed_form.txt};

\addlegendentry{Integral}
\addplot[Set1-C,integral,dashed] table[x=wav,y=bw1] {data/integral.txt};
\addplot[Set1-C,integral, dashed,forget plot] table[x=wav,y=bw3] {data/integral.txt};
\addplot[Set1-C,integral, dashed,forget plot] table[x=wav,y=bw10] {data/integral.txt};

\addlegendentry{Simulation}
\addplot[Set1-A,simulation,dotted] table[x=wav,y=bw1] {data/SSFM.txt};
\addplot[Set1-A,simulation,dotted,forget plot] table[x=wav,y=bw3] {data/SSFM.txt};
\addplot[Set1-A,simulation,dotted,forget plot] table[x=wav,y=bw10] {data/SSFM.txt};

\end{axis}

\end{tikzpicture}

\begin{tikzpicture}[baseline]
\begin{axis}[
unbounded coords=jump,
    title = (c) Forward + Backward Raman Amplification,
    legend style={font=\footnotesize, at={(rel axis cs:0,0)}, anchor=south west},
    width=\linewidth, height = 6cm,
    xlabel={Wavelength [nm]},
    ylabel={$\text{SNR}_{\text{NLI}}$ [dB]},
    grid=both,
    xtick={1490,1510,1530,1550,1570,1590,1610},
ytick={18,22,26,30,34,38,42},
    ymax=42,ymin=18,
    xmin=1490,xmax=1610,
    xticklabel style={/pgf/number format/1000 sep=},
    /pgf/number format/fixed,
    ylabel near ticks,
    ytick distance=4,
    xtick distance=20,
]
\begin{scope}[on background layer]
    \fill[green,opacity=0.1] ({rel axis cs:0.0,0.0}) rectangle ({rel axis cs:0.3,1});
    \fill[blue,opacity=0.1] ({rel axis cs:0.3,0}) 
    rectangle ({rel axis cs:0.65,1});
    \fill[gray,opacity=0.1] ({rel axis cs:0.65,0})
    rectangle ({rel axis cs:1,1});
\end{scope}
    
\node[anchor=west] (source) at (axis cs:1510,40.5){S};
\node[anchor=west] (source) at (axis cs:1545,40.5){C};
\node[anchor=west] (source) at (axis cs:1580,40.5){L};

\draw [black] \pgfextra{
    \pgfpathellipse{\pgfplotspointaxisxy{1550}{35.2}}
    {\pgfplotspointaxisdirectionxy{5}{0}}
    {\pgfplotspointaxisdirectionxy{0}{2}}
}; 

\node[anchor=west] (source) at (axis cs:1551,35.8){};
\node (destination) at (axis cs:1563,36.1){};
\draw[->](source)--(destination);

\node[] at (axis cs: 1577,36.1) {Single span}; 

\draw [black] \pgfextra{
    \pgfpathellipse{\pgfplotspointaxisxy{1550}{30.1}}
    {\pgfplotspointaxisdirectionxy{5}{0}}
    {\pgfplotspointaxisdirectionxy{0}{2}}
}; 
\node[] at (axis cs: 1528,28.7) {3 spans}; 

\node[anchor=east] (source) at (axis cs:1550,29.2){};
\node (destination) at (axis cs:1536,28.5){};
\draw[->](source)--(destination);

\draw [black] \pgfextra{
    \pgfpathellipse{\pgfplotspointaxisxy{1550}{24.8}}
    {\pgfplotspointaxisdirectionxy{5}{0}}
    {\pgfplotspointaxisdirectionxy{0}{2}}
}; 
\node[] at (axis cs: 1542,20.7) {10 spans}; 

\node[anchor=east] (source) at (axis cs:1553,23.5){};
\node (destination) at (axis cs:1544,21){};
\draw[->](source)--(destination);

\addlegendentry{Closed-form}
\addplot[Set1-B,closedform] table[x=wav,y=fwbw1] {data/closed_form.txt};
\addplot[Set1-B,closedform,forget plot] table[x=wav,y=fwbw3] {data/closed_form.txt};
\addplot[Set1-B,closedform,forget plot] table[x=wav,y=fwbw10] {data/closed_form.txt};

\addlegendentry{Integral}
\addplot[Set1-C,integral,dashed] table[x=wav,y=fwbw1] {data/integral.txt};
\addplot[Set1-C,integral, dashed,forget plot] table[x=wav,y=fwbw3] {data/integral.txt};
\addplot[Set1-C,integral, dashed,forget plot] table[x=wav,y=fwbw10] {data/integral.txt};

\addlegendentry{Simulation}
\addplot[Set1-A,simulation,dotted] table[x=wav,y=fwbw1] {data/SSFM.txt};
\addplot[Set1-A,simulation,dotted,forget plot] table[x=wav,y=fwbw3] {data/SSFM.txt};
\addplot[Set1-A,simulation,dotted,forget plot] table[x=wav,y=fwbw10] {data/SSFM.txt};

\end{axis}

\end{tikzpicture}